\documentclass[11pt]{article}
\usepackage[margin=2cm]{geometry}
\usepackage[normalem]{ulem}
\usepackage{authblk}
\usepackage{amsmath, amssymb}
\usepackage{graphicx}
\usepackage{doi}
\usepackage{hyperref}
\usepackage{cite}
\usepackage{enumitem}

 \newcommand{\added}[1]{#1}
 \newcommand{\removed}[1]{}

 \def\markeupstatement{}
\begin{document}

\title{\added{Chaos-Generating Periodic Orbits of Topological Defects in Confined Active Nematics}}

\author[a]{Brandon Klein}
\author[a,b]{Alejandro J. Soto Franco}
\author[c]{Md Mainul Hasan Sabbir}
\author[d]{Matthew J.~Deutsch}
\author[a]{Ross Kliegman}
\author[d,e]{Robin L.~B.~Selinger}
\author[c]{Kevin A. Mitchell}
\author[a]{Daniel A. Beller\footnote{To whom correspondence may be addressed. E-mail: d.a.beller@jhu.edu}}

\affil[a]{Department of Physics and Astronomy, Johns Hopkins University, Baltimore, Maryland 21218, USA}
\affil[b]{Department of Biomedical Engineering, Johns Hopkins University, Baltimore, Maryland 21218, USA}
\affil[c]{Department of Physics, University of California, Merced, California 95344, USA}
\affil[d]{Advanced Materials and Liquid Crystal Institute, Kent State University, Kent, Ohio 44242, USA}
\affil[e]{Physics Department, Kent State University, Kent, Ohio 44242, USA}

\def\authorcontributions{D.A.B., K.A.M., and R.L.B.S. supervised the project and, along with B.K., designed the study. B.K., R.K., M.J.D., R.L.B.S., and D.A.B.\ designed the numerical simulations. B.K., A.J.S.F., and M.J.D.\ performed the simulations. M.M.H.S.\ and K.A.M. designed and performed the numerical topological entropy analysis. B.K.\ conducted the analytical calculations. 
B.K.\ and D.A.B.\ wrote the paper with input from all authors.}

\maketitle

\begin{abstract}\markeupstatement
Active nematic\added{s} \removed{flows} in two dimensions\removed{, largely driven by motile +1/2 disclinations,} \added{stir} \removed{mix} themselves efficiently \removed{and exhibit chaos in the bulk steady state} \added{through internally generated chaotic flows, largely driven by motile $+1/2$ disclinations}. \removed{Motivated by recent experimental findings for three-defect braiding in cardioid-shaped domains, w}\added{W}e investigate how this tendency toward chaotic fluid \added{stirring}\removed{mixing} can, counterintuitively, produce certain ordered, periodic flows in confinement\added{, characterized by stable periodic orbits of $+1/2$ disclinations}\removed{ with a controllable net topological charge}. We \added{computationally} study two-dimensional active nematics in systems with boundary conditions requiring a prescribed number \added{$n$} of excess \removed{+1/2}\added{$+1/2$} disclinations, using Beris-Edwards nematohydrodynamics simulations\removed{,} alongside an agent-based\removed{, hydrodynamic} simulation approach. We find \added{that when confinement is sufficiently strong to prevent defect pair-nucleation, but not strong enough to arrest all flow, then $n=3$ defects generically follow a "golden braid" orbit as observed recently in experiments, and we predict a "silver braid" orbit of $n=4$ defects.  For these results and for greater numbers of defects, we show that the periodic or chaotic nature of the dynamics is determined by a balance between the number of defects and the number of vortices in the flow field, suggesting a new design criterion for ordered flows in active nematics.} \removed{ordered flows for systems of three and four defects, and we use tools from braid theory to show that spontaneously occurring periodic defect motions produce maximal topological entropy. Our theory correctly predicts the generic absence of stable periodic orbits of more than four defects in strong confinement in simulation. Our results identify the parameter regime outside of which periodicity is lost, and allow us to probe the limits of topological entropy production.}
\end{abstract}

Active systems\removed{, like all life, are} \added{exist} far from equilibrium, transforming energy from ambient or internal sources into a wide variety of complex collective motions, dependent on the interactions between the active particles \cite{bowick2022symmetry,Ramaswamy2017,RevModPhys.85.1143}. \added{In active nematics, apolar orientational order emerges but is interrupted by topological defects known as disclinations, characterized by a half-integer winding number \cite{doostmohammadi_active_2018,Giomi2014,sanchez2012spontaneous,decamp2015orientational,duclos2020topological}. Disclinations play important and diverse roles in the dynamics of many active nematic systems, including effects on cell density, cell death, and transitions from two- to three-dimensional behavior \cite{maroudas-sacks_topological_2021, copenhagen_topological_2021, duclos_topological_2017, hirst_liquid_2017, armengol-collado_epithelia_2023, hokmabad_topological_2019, shimaya_tilt-induced_2022}.}

\begin{figure*}[h!]
    \centering
    \includegraphics[width=\textwidth]{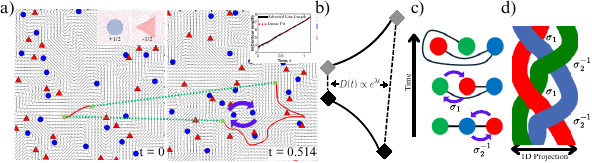}
    \caption{
    (a) Exponential stretching of a passively advected line, as a measure of topological entropy production, in simulated bulk active nematic dynamics beginning along an arbitrary contour of the nematic director field. Left inset shows illustrations of the local director field around $\pm 1/2$ topological defects. Right inset shows the semi-log plot of the advected contour length over time. \added{Green dotted lines indicate the divergence of the contour length between the two dotted green points. Purple arrows show the recent swapping of two $+1/2$ defects.}
    (b) Schematic illustration of the divergence of two passive tracers whose separation distance $D$ grows exponentially in time $t$, giving a positive Lyapunov exponent $\lambda$. (c) Schematic mixing dynamics for ``stirring rod'' motion described by the braid\removed{ }word $\{\sigma_{\removed{1}\added{2}}^{-1} \sigma_{\removed{2}\added{1}}\}$. The exponential stretching is deducible from the growth in \removed{perimeter}\added{length} of the black \removed{region}\added{line}. 
    (d) Worldlines of topological defects for the braidword shown in (c). 
    }
    \label{fig:1}
\end{figure*}

\added{Alongside their effects on living systems, disclinations have drawn great interest because they are central to the unusual dynamics characteristic of active nematics. The bulk steady state  features continual production and annihilation of disclinations in pairs of winding numbers $+1/2$ and $-1/2$ (see Fig.~\ref{fig:1}a inset) \cite{sanchez2012spontaneous,decamp2015orientational,giomi_geometry_2015}. While the $-1/2$ defects are passively advected, the $+1/2$ defects move persistently like self-propelled particles. 
The prediction \cite{marenduzzo2007steady} of such a topologically active steady state in hydrodynamic theory is borne out especially clearly in experimental model systems involving aqueous suspensions of cytoskeletal filaments (microtubules, f-actin) along with molecular motors (kinesin, myosin) \cite{sanchez2012spontaneous, kumar_2018_tunable}; the motors consume ATP and exert non-equilibrium forces that produce extensile active stresses.}

\added{In the microtubule-kinesin model system, it has been shown that self-propelled $+1/2$ defects act like stirring rods upon the surrounding fluid, dominating its self-stirring dynamics \cite{tan_topological_2019}. Without strong confinement, the apparently disordered motions of these defects cause the active nematic to stir itself chaotically, an unusual property for a system in the low-Reynolds number regime \cite{tan_topological_2019}.  The focus of this work is on ordered motions of defects that emerge spontaneously in strong confinement, which nonetheless drive chaotic self-stirring.}
\removed{Many such active states exhibit emergent orientational order, and systems that can be described as space-filling active fluids commonly feature excitations of the orientational order known as topological defects. Strikingly, topological defects have been found to play important and diverse roles in the dynamics of many active systems, both natural and synthetic, \cite{maroudas-sacks_topological_2021, copenhagen_topological_2021, duclos_topological_2017, hirst_liquid_2017, armengol-collado_epithelia_2023, hokmabad_topological_2019, shimaya_tilt-induced_2022}.}

\removed{This has inspired much}\added{Because of the central role of $+1/2$ disclinations in the internally generated flows of active nematics, there has been great} interest in describing and controlling \added{these }defects and their dynamics\added{, both} for fundamental understanding and for possible microfluidic and industrial applications. 
\removed{To these ends,  model active nematic systems have been constructed \textit{in vitro} from biological components. One common method\cite{inaba_assembling_2022, doostmohammadi_active_2018} involves 
microtubule bundles suspended in a fluid with ATP-powered kinesin dimers, which walk along the microtubules, pushing them in opposite directions and thereby injecting extensile activity into the system. 
With model systems such as these, much has been learned about defect control \cite{cui_effects_2017, doostmohammadi_stabilization_2016,hardouin_reconfigurable_2019, serra_defect-mediated_2023} and ordered flows \cite{calderer_chevron_2024, memarian_active_2021}.}%
\removed{Bulk active nematics are known to exhibit chaotic defect motion, and much work has been done in creating control mechanisms for their flow structures and dynamics to tame this chaos.}%
For example, \added{recent investigations have shown that }Gaussian curvature \removed{tends to charge separate}\added{promotes charge-separation of} topological defects \removed{with the same sign}\added{by the sign of their winding numbers} \cite{ellis2018curvature}\removed{,}\added{;} substrate friction \removed{has been shown to }tune\added{s} the characteristic length scales of director distortions in active nematics \cite{PhysRevE.90.062307}\removed{,}\added{;} and topographical patterns in the substrates underlying nematic fluids exhibit remarkable control over the rheological properties of the bulk \cite{thijssen2021submersed}. \removed{The realm of defect s}\added{S}patiotemporal control \added{of defects }has also been explored using activity gradients as effective \removed{electric}\added{external} fields for defect quasi-particles \cite{PhysRevX.9.041047, zhang2021spatiotemporal}, and has been achieved experimentally with light activation of myosin molecular motors \cite{zarei2023light}.

\added{Geometrically structured confinement, the control mechanism of interest in this work, has shown great promise for accessing distinct behaviors in which the apparently erratic defect trajectories of bulk active nematics are replaced by certain ordered, predictable motions.}\removed{One vital control mechanism for nematic flow structures is that of boundary conditions and the constraints they impose on bulk systems.} For example, when active nematics are confined to a disk \cite{opathalage_self-organized_2019}, an ordered state is produced in which two \added{defects of winding number}\removed{positive} $+1/2$ \removed{defects} circle around a common vortex core, periodically interrupted by $\pm 1/2$ defect pair nucleation and pair annihilation. 
In simulations of annuli and disks\removed{ with variable winding number}, defects \removed{have a tendency}\added{tend} to self-screen excess topological charge \cite{norton2018insensitivity, hardouin_active_2022} by localizing near boundaries. \removed{This effect is prominent when active nematics are confined to a channel, where they}\added{When confined to a channel of appropriate dimensions, active nematics} can produce an array of vortices around which defects \removed{rotate in an alternating} ``dance'' \added{along paths of alternating curvature} \cite{shendruk_dancing_2017}. \removed{The negative defects pin to the walls of the channel allowing for the smooth motion of $+1/2$ defects without annihilation. }When \removed{periodic }obstacles are placed \added{periodically} through\added{out} the nematic bulk, a 2D vortex lattice can be stabilized \cite{PhysRevLett.132.018301}. \removed{Both of these systems take place on periodic boundaries. }\removed{It was recently shown that a periodic 2D plane}\added{Periodic boundary conditions} alone can produce an ordered\added{, periodic} \removed{braiding }motion of $+1/2$ defects\added{, as can confinement to the surface of sphere \cite{keber2014topology,smith_braiding_2022}.}\removed{ around quasi-stationary negative defects \cite{mitchell_maximally_2024}. This braiding motion was shown to be identical to the braid conjectured, with strong numerical evidence, to be the maximally mixing braid on a topological annulus as measured by \emph{topological entropy} \cite{smith_topological_2022}.} 

\removed{Tan and coauthors \cite{tan_topological_2019} showed that the fluid mixing capabilities of active nematics are deducible from the braiding motions of the worldlines of +1/2 topological defects. The self propulsion of the +1/2 defects drives the dynamic evolution of the Lagrangian coherent structures which connect initially infinitesimally close passive tracers in the bulk, and is thus correlated to the production of a positive Lyapunov exponent (Figure \ref{fig:1}b).}

\added{Inspiring the present work,}\removed{Recently,} Memarian and coauthors \cite{mitchell_cardioid} \added{ recently used}\removed{showed that} a cardioid-like boundary \added{geometry to}\removed{can} pin a \added{$-1/2$}\removed{negative} defect at the \removed{inward-facing} \added{cardioid's} cusp, \added{thereby producing a closed geometry with a net topological charge of $3/2$ in the interior, rather than two as in disk-shaped confinement. When the system size was small enough that the only defects in the interior were the three topologically required $+1/2$ defects, but still large enough that these defects were mobile, 
 all three positive defects spontaneously adopted the same periodic orbit,}\removed{allowing three motile positive defects to encircle each other in alternating swaps,} in a pattern known as the golden braid. \added{Intriguingly, t}\removed{T}his braid is \removed{proven to be }the  optimal mixing \removed{braid}\added{motion} for three ``stirring rods''\added{, in a specific sense defined below}  \cite{finn_topological_2011}. 

These findings highlight a need for more general understanding of \added{periodic defect motions in confined active nematics: what types of periodic motions can arise spontaneously, which boundary geometries enable them, and how do they affect the active self-stirring?}\removed{ how topological defect dynamics in active nematics respond to the geometry and topology of confining boundaries.} In this work, we \added{computationally }investigate governing principles responsible for the golden braid in confined active nematics with three $+1/2$ defects, and \added{we explore }whether similar periodic orbits can be obtained with \removed{other}\added{greater} numbers of motile defects. \added{We primarily use simulated Beris-Edwards nematohydrodynamics, with corroborating evidence from a new, agent-based model of active  filaments coupled to a coarse-grained fluid.}

\added{First, to explore the generality of the golden braid result, we study a hypothetical system in which we can freely tune the net topological charge, and thus the number $n$ of excess $+1/2$ defects, in an active nematic confined to a disk. This allows us to separate the topological effects of the boundary as a whole from the localized influence of the cardioid's cusp. We show that the golden braid arises spontaneously in this system for $n=3$. We then observe a new periodic orbit for $n=4$, the silver braid,  and an absence of periodic orbits in favor of erratic trajectories for $n\geq 5$.
Next, we demonstrate that the variation of active force at the boundaries has a topological character that allows us to rationalize the commonalities of defect orbits in our hypothetical disks with those of the cardioid-confined system with tangential anchoring, and its experimentally realistic generalizations. We propose a general principle linking our findings for different $n$, which establishes a connection between braiding defect trajectories and the topology of vortices in the flow field. 
Finally, we numerically test the predictions arising from our hypothetical boundaries by simulating  active nematics confined inside cardioids, or similar curves having more than one cusp, with experimentally realistic tangential anchoring. In the parameter regime featuring mobile $+1/2$ defects and no other defects than the topologically required ones, these simulations reproduce our predicted periodic braids or lack thereof, with one notable exception that underscores the connection between defect braids and the flow field's vortex structure.}

\section*{Results}

\added{Our computational approach simulates the time-evolution of the flow velocity field $\mathbf{u}(\mathbf{r})$ along with the Q-tensor $Q_{ij}(\mathbf{r}) = S(\mathbf{r}) (n_i(\mathbf{r}) n_j (\mathbf{r}) - \tfrac{1}{2} \delta_{ij})$. Here, $S$ is the scalar degree of nematic order with an arbitrarily chosen global rescaling,  which for convenience we set to give a bulk equilibrium value $S_0 = \sqrt{2}$; $\mathbf{n}$ is the nematic director; $\delta_{ij}$ is the Kronecker delta; and $i,j$ both run over $x,y$. We employ a strong anchoring limit under which $Q$ is held fixed at the boundaries, meaning that we hold constant $\mathbf{n}$ as well as $S$ there.}

\added{We simulate active nematic dynamics using active Beris-Edwards nematohydrodynamics \cite{beris1994thermodynamics} (equations of motion given in Materials and Methods), a widely used approach \cite{marenduzzo2007steady,doostmohammadi_active_2018}. We initialize our system with the equilibrium scalar degree of order, $S$, and a random director field at every point in space. In this framework there are two characteristic length scales: The first is the active length scale $\ell_a = \sqrt{K/\zeta}$ which sets the typical spacing between defects produced by activity (rather than boundary topology); here $K$ is the single Frank elastic constant for splay and bend distortions and $\zeta$ is the activity coefficient determining the strength of the extensile active stress. The second length scale is the nematic coherence length $\ell_c$, which sets the characteristic size of the defect core within which nematic order $S$ is diminished. We vary these length scales to explore the resulting active steady states. 
We report nondimensionalized values of $\ell_c$ and $\ell_a$ obtained by dividing both these lengths by the effective system length, given by the square root of the system area in units of lattice spaces: $\tilde \ell_a = \ell_a / \sqrt{A_{\text{sys}}}$, $\tilde \ell_c = \ell_c / \sqrt{A_{\text{sys}}}$. The tilde notation is omitted for brevity. We employ a purely finite difference scheme for updating both $\mathbf{u}$ and $Q$.} 

\added{Chaos in the flow field is characterized by an exponential growth of the distance $D$ between initially nearby points, with exponential rate $\lambda$ called the Lyapunov exponent (Fig.~\ref{fig:1}b). To quantify chaos for the periodic defect motions considered in this work, we measure a closely related quantity, the so-called topological entropy $h$, which is the exponential rate of stretching measured in the length $L$ of a curve of tracers passively advected in the flow (Fig.~\ref{fig:1}a).  The topological entropy is an upper bound on $\lambda$ \cite{thiffeault_topology_2006}  and in experiments  on microtubule-based active nematics, this bound is nearly met \cite{tan_topological_2019}. Also, it is possible to analytically calculate topological entropy in periodic flows (see Supporting Information) \cite{thiffeault_topology_2006}. Therefore, we use $h$ as a proxy for $\lambda$, meaning that we measure the chaotic character of active nematics by measuring the stretching injected into the bulk fluid.  As shown in Ref.~\cite{tan_topological_2019}, $h$ is dominated in active nematics by the motions of $+1/2$ defects, which push advected tracer curves ahead of them. Therefore, if there are $n$ positive defects that never annihilate with a negative defect, $h$ is the same exponential growth rate as that of a curve that goes through all the defects exactly once, always in the same order, even as the defects interchange positions. The increasing length $L(t)$ of such a curve is illustrated schematically in Fig.~\ref{fig:1}c.} 

\added{Importantly, the growth of $L$ is exponential only for certain motions of the $+1/2$ defects, determined by the way in which their trajectories braid around one another. Fig.~\ref{fig:1}c illustrates a case of exponential growth for $n=3$. First, the rightmost two defects swap their positions in a counterclockwise (CCW) manner. Second, the leftmost two defects swap their positions in a clockwise (CW) manner. If these two swaps repeat in alternation, then $L(t)$ grows exponentially in time with each period $T$ as $e^{h_T t/T}$, where $h_T = 2\log \phi_0$, with $\phi_0 = (1 + \sqrt 5)/2$ being the golden ratio \cite{thiffeault_measuring_2005}. It is no accident that the minimal length of the curve if pulled taught (the ``train tracks" construction \cite{Bestvina_1992}) each step grows according to the Fibonacci sequence, from $2$ to $3$ and then $5$, followed by $8$ (and so on) if the cycle is repeated. Figure~\ref{fig:1}c in fact depicts the topology of the golden braid, named for its connection to the golden ratio, as observed experimentally in the cardioid-confined active nematic of Ref.~\cite{mitchell_cardioid}, where the measured $h_T$ agreed well with this prediction. However, if the two swaps occur in the same sense (both CW or both CCW) then the growth of $L(t)$ is linear, not exponential, in time.}

\added{To label the swaps and record their handedness, we adopt the language of braid generators: a swap of the $i$th and $(i+1)$th defects is labeled $\sigma_i$ if it occurs CW, or $\sigma_i^{-1}$ if it is CCW \cite{thiffeault_measuring_2005}. A periodic motion of defects can then be described by a braidword, which is an ordered sequence of braid generators describing the motion in one period. For the golden braid of Fig.~\ref{fig:1}c, the braidword is $\{\sigma_2^{-1} \sigma_1\}$. In order for this language of braid theory to be useful for defects that are able to move about a two-dimensional domain, we must have an unambiguous way to order them. For this purpose we choose the simple method of projecting positions onto one spatial axis $x$, so that a ``swap'' is recorded each time the $x$-ordering changes, and we record the handedness of the swap as CW if the defect with lower final $x$-value appears to pass in front of the other defect, and CCW otherwise (Fig.~\ref{fig:1}d). The spatially projected defect trajectories plotted along one spatial and one temporal axis, maintaining their crossing order, produces a braid of $n$ strands, as depicted for the golden braid in Fig.~\ref{fig:1}d.}

\added{This notation from braid theory along with spatial projection enables us to characterize the topological properties of periodic defect motions in confined active nematics, which are comparable across different confinement geometries. Furthermore, because there is an established method to compute the topological entropy $h$ of any braidword (see Supporting Information) \cite{thiffeault_measuring_2005,thiffeault_topology_2006,smith_braiding_2022,smith_topological_2022}, we have quantitative predictions of the chaos produced by any periodic defect motion.}
 
\subsection*{Circular confinement with controllable topological charge}

 \removed{By viewing +1/2 defects as self-propelled stirring rods, we can measure the amount of stretching their dynamics inject into the bulk and quantitatively distinguish dynamic patterns by their fluid mixing capability. Thus, controlling the number of motile charges is key to controlling topological mixing.
 To systematically tune the number of mobile $+1/2$ defects, we consider circular confinement of the active nematic phase while varying the anchoring direction of the nematic director on the boundary. In particular, with the circular boundary $\partial \Omega$ parameterized by angle  $\theta$, we initialize the nematic order as
\begin{equation} 
     {\bf n}\big\rvert_{\partial \Omega} 
     (\theta) =
     \pm \left(\begin{array}{r}
        -\sin(q\theta)\\ 
        \cos(q\theta)
    \end{array}\right), \nonumber
\end{equation}
and set the scalar degree of order, $S$, to $\sqrt 2$. 
The $Q$-tensor, defined by Equation ~\ref{eq:Q_def}, is held fixed at the boundaries.
This boundary condition produces a net topological charge $q$ in the nematic domain $\Omega$, causing the lowest-energy states to have  $n=2q$ topological defects with winding number $+1/2$.}
\added{We begin by constructing a hypothetical confinement scenario in which the boundary shape is a circle, and thus smooth, but the total winding number $q$ of the enclosed defects can be tuned to any half-integer. To achieve this, we impose a spatially varying anchoring direction as a function of azimuthal coordinate $\theta$. In particular, we choose:
\begin{equation} \label{eq:boundary_Q}
     {\bf n}(\theta)\big\rvert_{\mathrm{bdy}}  =
     \pm \left(\begin{array}{r}
        -\sin(q\theta)\\ 
        \cos(q\theta)
    \end{array}\right).
\end{equation}
The steady-winding circle boundary defined by Eq.~\ref{eq:boundary_Q} ensures a total winding $2\pi q$ in the directory at the boundary, requiring the net topological charge of defects in the interior of the active nematic to be $q$. This topological requirement is met most simply, and at lowest free energy, by $n=2q$  defects of $+1/2$ winding and no $-1/2$ defects. We examine values of $\ell_a$ for which the $+1/2$ defects are motile but there is no defect pair-nucleation or annihilation, so that the same $n$ positive defects exist at all times after an initial transient. Except for $q=1$, Eq.~\ref{eq:boundary_Q} is not realistic for experiments, where anchoring is generally tangential. Our motivations to analyze this scenario first are its simplicity and analytical tractability, and because it separates the effects of topology from the sharpness of boundary geometries, which are inextricably linked in the cardioid. Later, we show that features of this hypothetical system are topologically robust and applicable to experimentally realistic scenarios.}

We first consider \added{the case $q=1$, which is simply} a circular boundary with tangential anchoring\added{, with $\ell_a$ sufficiently large as to prevent pair creation}\removed{, providing a net charge of $q=+1$, thus requiring two $+1/2$ topological defects}. 
\begin{figure*}[h!]
    \centering
    \includegraphics[width=\textwidth]{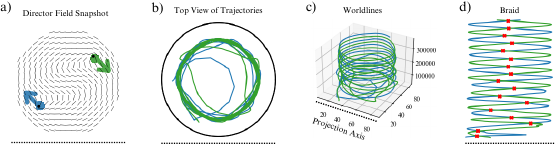}
    \caption{Braiding terminology demonstrated with two $+1/2$ defects in an active nematic confined in a circle with tangential anchoring, from a $100 \times 100$ Beris-Edwards nematohydrodynamic simulation. \added{(a) Snapshot of the nematic director showing the two defects in a periodic orbit. }(\removed{a}\added{b}) Trajectory \removed{traces}\added{top view} of defects (blue and green) over simulation time (335,000 time-steps). \removed{Inset shows a snapshot of defects during cyclic motion.} (\removed{b}\added{c}) Worldlines of the \removed{defects}\added{trajectories} in (\removed{a}\added{b). Spatial coordinates and time are labeled on the $x, y$, and $z$ axes respectively}. (\removed{c}\added{d}) Defect worldlines spatially projected onto the $x$ axis. Crossings (all clockwise) are labeled by a red ``x''; the structure is represented by the one-element braidword $\{\sigma_1\}$.}
    \label{fig:2}
\end{figure*}
This system was realized \added{experimentally }by Ref.~\cite{opathalage_self-organized_2019} and, similarly, we find that the two positive defects \removed{circle}\added{follow a circular path in} a common direction \added{after a brief transient interval lasting about one period}, 
\removed{creating the braidword $\{\sigma_1^{\pm 1}\}$ and zero  topological entropy,}%
as shown in Fig. \ref{fig:2}\added{a,b} and Movie S1. 
\removed{The braidword taxonomy and topological entropy calculations are described in the Materials and Methods.}
\added{We note that i}\removed{I}n the experimental system of Ref.~\cite{opathalage_self-organized_2019}, the co-rotating defects deviate from their quasi-circular trajectories upon the nucleation of a \added{$\pm 1/2$} defect pair from the boundary, whose negative defect annihilates one of the bulk defects. Beris-Edwards nematohydrodynamics, however, do not observe such breaking of periodicity below the activity threshold to active turbulence\added{, in agreement with previous simulation studies} \cite{schimming2023friction, mirantsev2021behavior, joshi2023disks}. 
\removed{Likewise, we find that the co-rotating defects are an active steady state.}
\added{With only two strands, the only braid generators are $\sigma_1$ and $\sigma_1^{-1}$, and no braidword can be constructed from these which produces a nonzero topological entropy.}
\removed{The only braidwords that can be created in $\mathbf{B}_2$ (i.e.,\ with two defects) are $\{\sigma_1^{\pm 1}\}$ and the identity, which both have a maximum eigenvalue of 1.}
This is consistent with the fact that two stirring rods are incapable of producing chaotic \added{stirring}\removed{mixing}~\cite{thiffeault_topology_2006}.

\begin{figure*}[h]
    \centering
    \includegraphics[width=\textwidth]{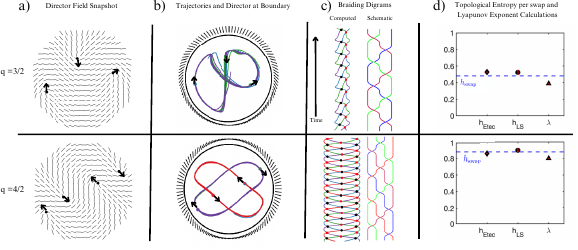}
    \caption{Periodic braiding orbits of $+1/2$ defects in a simulated active nematic confined in a disk with an anchoring direction that winds through angle $2\pi q$, for $q=3/2$ (top row) and $q=4/2$ (bottom row). \added{(a) Snapshot of the nematic director showing the three ($q=3/2$) and four ($q=4/2$) defects in a periodic orbit.} (\removed{a}\added{b}) The anchoring direction along the circular boundary and defect trajectories traced over the simulation time of $7.5\times10^5$ ($q=3/2$) and $5.0\times10^5$ ($q=4/2$) time-steps. Each simulation was performed on a $100\times100$ lattice with a dimensionless active length of $0.045$, and a dimensionless nematic coherence length of $0.011$. Arrows indicate defect direction of motion. (\removed{b}\added{c}) (left) The projection of the trajectories onto the $x$ axis, where swaps between defects are labeled with a red \removed{``x''}\added{marker} if clockwise and a black \removed{dot}\added{marker} if counter-clockwise; (right) schematic diagram summarizing the braid exhibited by the defects \removed{$\{\sigma_1\sigma_2^{-1}\}$ }\added{$\{\sigma_2^{-1}\sigma_1\}$ } for $q=3/2$ and $\{\sigma_1 \sigma_3 \sigma_2 \sigma_1^{-1} \sigma_3^{-1} \sigma_2^{-1}\}$ for $q=4/2$. (\removed{c}\added{d}) The numerically calculated topological entropy using the E-tec and Line Stretching (LS) algorithms (shown in Movies S2 and S3), as well as the calculated Lyapunov exponent, in units reciprocal to the time between defect swaps. Each braiding pattern consists of two effective swaps corresponding to a co-linear arrangement of defects. The numerical values of average topological entropy per swap using E-tec are $0.5277 \pm 0.0005$ ($q=3/2$) and $0.8610 \pm 0.0004$ ($q=4/2$) for samples of 3000 randomly initialized advected trajectories. The numerical values of average topological entropy per swap using the LS algorithm are $0.52 \pm 0.03$ ($q=3/2$) and $0.90 \pm 0.02$ ($q=4/2$). Errors are standard error of the mean taken over five advected curves. The numerical values of average Lyapunov exponent per swap are $0.3873 \pm .0003$ ($q=3/2$) and $.8006 \pm .0004$ ($q=4/2$)\removed{. Each uses 350 pairs of randomly initialized passive tracers. B}\added{,b}oth of which are, as required, below their respective analytic values of topological entropy for ideal stirring rods, shown in the dashed blue line. \added{Each $\lambda$ calculation uses 350 pairs of randomly initialized passive tracers. } \removed{Error bars are}\added{Standard deviation is} shown in red and are smaller than the marker size.}
    \label{fig:3}

\end{figure*}

For $q=3/2$, we observe that the three $+1/2$ defects spontaneously produce the ``golden braid'' \added{as in Fig.~\ref{fig:1}c,d and experimentally }observed in cardioid-shaped confinement in Ref.~\cite{mitchell_cardioid}. \added{All three defects follow a single path with ``figure 8'' topology. From the perspective of any one traveling defect, the trajectory turns alternately clockwise and counterclockwise.} %
\added{With the trajectories projected onto any one spatial axis, }\removed{C}\added{c}onsecutive \removed{inversion events}\added{swaps} in the golden braid alternate between CW and CCW, and the same two defects are never swapped before one of them is swapped with the third. Traces of the defect trajectories, along with the \removed{worldlines}\added{braiding diagrams} of their $x$-projections, are shown in the top row of Fig.~\ref{fig:3}\removed{a, b}\added{b, c} and in Movie S2\added{, confirming that the braidword is $\{\sigma_2^{-1}\sigma_1 \}$}. 
\removed{The golden braid is described by the braidword $\{\sigma_1\sigma_2^{-1}\}$, which yields a topological entropy (per cycle) of $h=2\log \phi_0$, where $\phi_0 = (1 + \sqrt 5)/2$ is the golden ratio, for which the golden braid is named~\cite{thiffeault_measuring_2005}.} 
\added{The theoretically predicted topological entropy per period is therefore $h_T = 2 \log \phi_0$ as mentioned previously. }We normalize this by the number of swaps per cycle, \removed{$h_{\mathrm{swap}} = h / N_{\mathrm{swaps}}$}\added{$h_{\mathrm{swap}} = h_T / N_{\mathrm{swaps}}$}, with $N_{\mathrm{swaps}}=2$ in this case\removed{, where a swap corresponds to a co-linear arrangement of defects during the trajectory}. \added{In Fig.~\ref{fig:3}d, we show that t}\removed{T}his prediction agrees well with the topological entropy per swap as numerically calculated using \added{two established methods, }\removed{both} the Line Stretching and E-tec schemes\added{, as well as the calculated lower bound of the Lyapunov exponent. Methods used to compute all three of these measures are described in Materials and Methods}\removed{ (Fig.~\ref{fig:3}c)}.
 
\added{Turning now to the steady-winding circle}\removed{The} boundary condition \removed{producing}\added{with} $q=4/2$\added{, we obtain}\removed{results in} a \added{distinct} periodic active steady state of four $+1/2$ defects, which to our knowledge has not previously been studied in an active nematic. Defect trajectories and projection \removed{worldlines}\added{braid diagrams} are shown in the bottom row of Fig.~\ref{fig:3}\removed{a, b}\added{b, c} and in Movie S3. The braidword for this motion is $\{\sigma_1 \sigma_3 \sigma_2 \sigma_1^{-1} \sigma_3^{-1} \sigma_2^{-1}\}$\added{.}\removed{ = $\{\beta_{\mathrm{max},4}\beta^*_{\mathrm{max},4}\}$, where $[\cdot]^*$ denotes an element-wise inverse, and $\beta_{\mathrm{max},4}$}
\added{Interestingly, the braidword $\{\sigma_1\sigma_3 \sigma_2\}$, which is repeated alternately CW and CCW by the four defects,}  is conjectured by Finn and Thiffeault  \cite{finn_topological_2011} to be the braid of four stirring rods which maximizes topological entropy per swap\added{, with strong numerical evidence in favor of this conjecture}. Because the associated topological entropy \removed{per period is $h=2\log(\phi_1)$}\added{per period is $h_T=2\log(\phi_1)$}, where $\phi_1 = 1 + \sqrt 2$ is \added{called }the silver ratio, this braid is known as the ``silver braid'' \cite{thiffeault_measuring_2005}. 

Unlike the $q=3/2$ case, here the \added{four} $+1/2$ defects do not all follow the same trajectory; instead, there are two intersecting, mirror-image trajectories, each containing two of the four defects. 
\removed{
This braid is topologically equivalent to the ``Ceilidh dance''~\cite{shendruk_dancing_2017} with four $+1/2$ defects, with the distinction that the topology of our system is a disk, not an annulus, and therefore lacks the pinned negative defects seen in annular confinement.
}
\added{Arbitrarily choosing to order the defects by horizontal (or vertical) position at the time of the snapshot in Fig.~\ref{fig:1}a bottom panel, the silver braid is simply described by two ``physical swaps'' repeating in alternation: a CW swap of defects 1 and 4 and a CCW swap of defects 2 and 3. 
Thus, in Fig.~\ref{fig:3}d bottom panel, we present the theoretical and measured topological entropy with the normalization of $h$ per physical swap operation, $h_{\mathrm{swap}} = h_T / N_{\mathrm{swaps}}$, where $N_{\mathrm{swaps}}=2$.
}
\removed{Since all four defects are co-linear twice during each periodic cycle throughout the trajectory of this braid, we normalize the topological entropy per cycle by $N_{\mathrm{swaps}}=2$ swaps to obtain $h_{\mathrm{swap}}$.}

\begin{figure*}[h!]
    \centering
    \includegraphics[width=\linewidth]{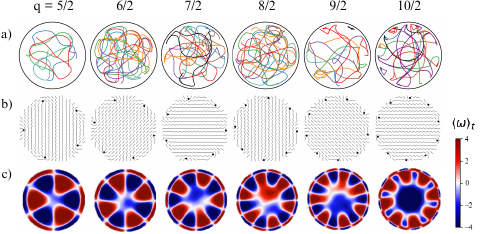}
    \caption{Defect dynamics in circular confinement with winding number $q\geq 5/2$ in the anchoring direction.  Columns show the different studied values of $q$. (a) \removed{S}\added{Ex}amples of the aperiodic trajectories. For the $q = 5/2$ system, trajectories over 40 time-steps are shown. For the $q = 6/2,7/2,8/2,9/2$ systems, trajectories are shown over 100 time-steps and for $q = 10/2$  300 time-steps are shown. (b) The passive ground state configurations of the defects. In all studied geometries, the ground states show a symmetric placement of defect cores about the boundary. However, orientations of the defects are not symmetric, with orientations varying locally to match the fixed anchoring against the circular boundaries. (c) The time averaged vorticity shown throughout the simulation time of $10^6$ time-steps. Each simulation was performed on a $100\times100$ lattice, at a dimensionless active length of .0003, and a dimensionless nematic coherence length of 0.011.}
    \label{fig:4}
\end{figure*}
In contrast to the systems of two, three, and four defects\added{, where we find periodic dynamics for a broad range of $\ell_a$ and $\ell_c$}, \removed{we find that }\added{steady-winding circle }boundary conditions with $q$ between $5/2$ and $10/2$ \added{(highest tested) }do not spontaneously adopt periodic motions of their five to ten $+1/2$ defects, as seen in ~\ref{fig:4}a\removed{.}\added{; instead, the defects move erratically. This distinction of periodic motion for $q=3/2$ or $4/2$ but aperiodic motion for $q\geq 5/2$ remained true over the entire parameter regime tested with varying $\ell_a\added{\in[2,10]/88.6}$ and $\ell_c\added{\in[1,5]/88.6}$, where $\sqrt{A_{sys}} \approx 88.6$ lattice points.  To rule out artifacts of our random initialization, we ran 50-trial ensembles for each of $q=3/2$, $4/2$, and $5/2$\added{ within this parameter range}, obtaining the same qualitative result in all runs. For comparison, t}\removed{T}he ground states \removed{for}\added{of} a passive nematic in the same geometries are shown in Fig.~\ref{fig:4}b, revealing $n$-fold rotational symmetry in the equilibrium locations of \added{the} $n$\removed{, $+1/2$}\added{ positive} defects\added{,} close to the boundary.  \removed{Across all studied active length scales $\ell_a$ and nematic coherence lengths $\ell_c$, these systems do not exhibit a periodic motion of defects as their active steady state.} \removed{For different random initializations of the bulk director, these systems tend to start in motion consistent with the braidword
\begin{equation}
    \{\sigma_n \sigma_{n-2}...\sigma_{n - 1}\sigma_{n-3}...\sigma_n^{-1} \sigma_{n-2}^{-1}...\sigma_{n - 1}^{-1}\sigma_{n-3}^{-1}...\}. \label{eq:general_braidword}
\end{equation}
However, adherence to this braid never exceeds one cycle and is replaced by apparently random motions partway through the braidword.}

\removed{
Despite the lack of periodic braiding in $n\geq 5$ defects, we observe a great deal of structure in the time-averaged vorticity\added{ of the flow field}, plotted in Fig.~\ref{fig:4}c. We will focus on the $n=5$ defect case when discussing unstable systems for the rest of this work, as our understanding of the instability of higher charge cases applies to all $n \geq 5$ systems.
}

It is interesting to note that $\ell_a$ has to be decreased as $q$ increases in order for the defects to remain mobile. As seen in Fig.~\ref{fig:4}a, some defects in the $q = 9/2$ and $q = 10/2$ systems remain close to their passive ground-state positions throughout the simulation. This is consistent with previous work showing that defects screen boundary charge \cite{norton2018insensitivity}, \added{albeit here with positive rather than negative defects}\removed{with the distinction that here we observe this effect in \emph{positively}-charged defects}. 

\added{
Despite the lack of periodic braiding for $q\geq 5/2$, we observe a great deal of structure in the time-averaged vorticity\added{ of the flow fields}, plotted in Fig.~\ref{fig:4}c. In particular, there are $4|q-1|$ regions of alternating vorticity. 
}
\subsection*{Boundary-imposed forces}
\begin{figure*}[h!]
    \centering
    \includegraphics[width=\textwidth]{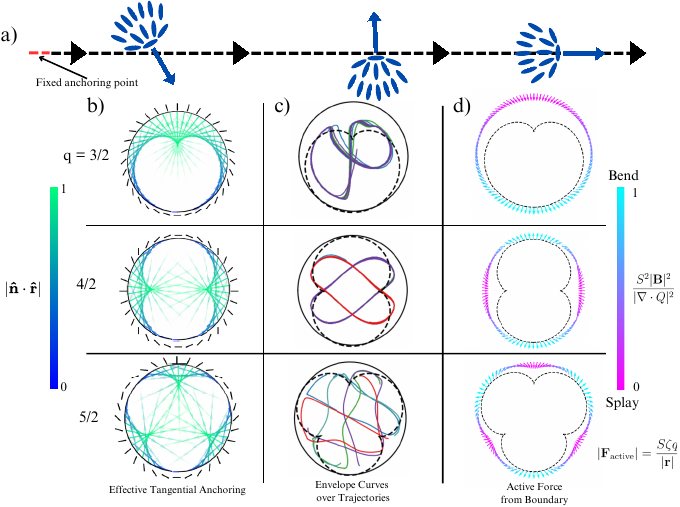}
    \caption{(a) Schematic illustration of a $+1/2$ defect in three orientations and positions consistent with a line of effective tangential anchoring; arrows mark defect self-propulsion direction. The only defect orientation \removed{whose trajectory}\added{which} does not break an anchoring line is \added{oriented }parallel to that line and \removed{which is} away from the fixed anchoring point on the boundary. \removed{Trajectories are colored by angle with respect to $\hat r$.} (b) The set of lines parallel to the anchoring direction \added{(anchoring lines)} at the associated point on the surface, showing the cardioid, nephroid, and trefoiloid \added{as emergent effective boundaries. Anchoring lines are colored by angle with respect to the radial direction $\hat r$ and have opacity decaying with distance from the anchoring point.} (c) Envelope curves (dashed) extracted from the lines of (b), together with simulated defect trajectories. Defect trajectories tend to stay approximately within the envelopes and to intersect with the cusps in the envelope. (d) The active force (Eq.~\ref{eq:f_active_epicycloid}) imposed by the winding anchoring conditions, with splay-dominated regions in pink and bend-dominated regions in cyan. In (b)-(d), each row corresponds to the anchoring winding $q$ labeled at left.}
    \label{fig:5}
\end{figure*}

 To understand the emergence of the observed defect braids for $n\leq 4$, as well as the generalizability of the dynamics they realize, we examine the relationship between defect trajectories and the forces and torques imposed by the boundary conditions. We observe that the tangent to a defect trajectory at a given point tends, when possible, to be parallel to a straight \removed{line}\added{ray} drawn through some point on the boundary and oriented along the director there. From Eq.~\ref{eq:boundary_Q}, these lines, \added{parametrized by $u\in [0, 2\pi)$,} are given by\added{ the $(x,y)$ values satisfying $f(u, x, y) = 0$ where}
\removed{
\begin{equation}
    \begin{aligned}
        f(u) = (y - r\sin(u)) + \cot(qu)(x - r\cos(u))= 0, \\u \in [0,2\pi).
    \end{aligned}
\end{equation}
}
\added{
\begin{align}
            f(u, x, y) = (y - r\sin(u)) + \cot(qu)(x - r\cos(u)).
    \label{eq:boundary_line}
\end{align}}
Here, $r$ is the radius of the circular confinement. This alignment can be understood geometrically as a dynamically stable situation for a self-propelling $+1/2$ defect. Since the defect experiences an active force in the direction of its comet-head, its tail is the only side that can be stably aligned with a fixed director at the boundary\added{ along a straight ray parallel to that director}. This idea is illustrated in Fig. ~\ref{fig:5}a\removed{, where defects with three distinct orientations are drawn along with a line on which the director is imagined to be held fixed parallel to the line. The only defect that is aligned with the fixed director more than instantaneously is the one whose self-propulsion lies along the line}. If we now consider the family of lines defined by Eq.~\ref{eq:boundary_line} (Fig.~\ref{fig:5}b), the scheme of orienting defects parallel to these \removed{lines}\added{rays} results in \removed{a dynamic stability for points in the interior region, whereas for certain regions near the boundary, many possible trajectories converge, pushing defects into the interior region}\added{defects moving away from parts of the boundary where the anchoring direction is more perpendicular}.

\added{In fact, there is a curve (dashed curves of Fig.~\ref{fig:5}c) along which a $+1/2$ defect traveling tangent to the curve would remain parallel to this boundary-imposed orientation at each point. This curve, the envelope of the family of rays, separates the region accessible to defects from a region or regions where we expect defects never to appear, to the extent that they follow our empirical boundary-alignment rule.}
\removed{The boundary between these two types of regions is the envelope of the family of lines, which is tangent at each point to one such line.}%
\added{This envelope is uniquely determined as the $(x,y)$ points satisfying both $f(u,x,y)=0$ and $\partial f/\partial u = 0$, with $f$ as given in Eq.~\ref{eq:boundary_line}. This gives}\removed{ The unique solution for this envelope satisfying $f(u) = \partial_u f(u) = 0$ is given by} 
\begin{align}
        x(u) = \frac{r}{2q}[(2q-1)\cos(u) + \cos((2q-1)u)] & \\
y(u) = \frac{r}{2q}[(2q-1)\sin(u) + \sin((2q-1)u)], &\;   u\in [0,2\pi).
\end{align}
These are the equations defining an epicycloid \cite{weisstein2003epicycloid}. Geometrically, epicycloids are  constructed by tracing the path of a point on a circle of radius $r_c$ as it rolls on the circumference of a circle $r$. In \removed{this}\added{our} construction, $r/r_c = 2(q - 1)=n-2$, which is the number of cusps in the resulting closed curve. For $q = 3/2$, $4/2$\added{,} and $5/2$, the envelopes are called a cardioid, a nephroid, and a trefoiloid respectively \cite{weisstein2003epicycloid}. 
\added{It is noteworthy that the shape of a cardioid emerges as a spontaneously obeyed boundary in a system constructed to resemble the cardioid-shaped confinement only topologically.}

We observe that these epicycloid envelopes predict certain important features of the simulated defect trajectories, as shown in \removed{Figure}\added{Fig.}~\ref{fig:5}c. 
Defect trajectories remain approximately inside the envelopes. Furthermore, over certain regions \removed{beginning}%
near a cusp in an envelope\removed{ and ending at that cusp}%
, the defect trajectories approximately coincide with the envelope while undergoing large-angle reorientation\added{, departing the envelope to return to the interior approximately when they reach the cusp}.  

We can understand how active forces yield this \removed{relationship}\added{behavior}  by \removed{identifying} \added{estimating the active force density at the boundary due to distortions imposed by the anchoring conditions Eq.~\ref{eq:boundary_Q}, ignoring radial variation in the director at the boundary.}
\def\factive{\mathbf{F}_{\mathrm{active}}}
\removed{
\begin{equation}
    \factive\vert_{\mathrm{bdy}} = -\zeta \nabla \cdot Q\vert_{\mathrm{bdy}} = \frac{S\zeta q}{|{\bf r}|} \begin{pmatrix}
        \cos((2q - 1)\theta) \\
        \sin((2q - 1)\theta)
    \end{pmatrix}.
\end{equation}
}
\added{
\begin{align}
     \factive\vert_{\mathrm{bdy}} = -\zeta \nabla \cdot Q\vert_{\mathrm{bdy}} = \frac{S\zeta q}{|{\bf r}|} \begin{pmatrix}
        \cos((2q - 1)\theta) \\
        \sin((2q - 1)\theta)
    \end{pmatrix}.
    \label{eq:f_active_epicycloid}
\end{align}
}
 It is useful to compare the right-hand side of Eq.~\ref{eq:f_active_epicycloid} to ${\factive = S\zeta({\bf S}-{\bf B})}$ where ${\bf S} = \mathbf n (\nabla \cdot \mathbf n )$ is the splay vector and ${\bf B} = \mathbf n \times (\nabla \times \mathbf n )$ is the bend vector. The bend contribution along the boundary has squared magnitude 
\removed{$$|{\bf B}|^2 = \frac{1}{S^2\zeta^2}|{\bf n} \cdot {\factive}|^2 = \frac{q^2}{r^2}\sin^2((q-1)\theta),$$}
\added{$$|{\bf B}|^2 = {S^{-2}\zeta^{-2}}|{\bf n} \times {\factive}|^2 = q^2 r^{-2} \sin^2((q-1)\theta),$$}
which is maximal at places where the anchoring is tangential, and the splay contribution has squared magnitude 
\removed{$$|{\bf S}|^2 = \frac{1}{S^2\zeta^2}|{\bf n} \times \factive|^2 = \frac{q^2}{r^2}\cos^2((q-1)\theta),$$}
\added{$$|{\bf S}|^2 = S^{-2}\zeta^{-2}|{\bf n} \cdot \factive|^2 = q^2 r^{-2} \cos^2((q-1)\theta),$$}
which is maximal at places where the anchoring is radial.
We find that the defects tend to move along tangentially anchored boundaries, where the bend mode dominates the active force, until the defect reaches the next radially-anchored location where the splay mode of the active force reorients the defect trajectory into the bulk. These splay maxima occur at the $\theta$-value of a cusp in the envelope curve, as each cusp lies along a line connecting the center of the system to a boundary point where the anchoring points radially along that line. Thus, defects experience maximum boundary-imposed splay at the envelope cusps, pushing the defects toward the center as we observe.  

The utility of the epicycloid envelopes is then two-fold: they can predict all of the locations of sharp reorientation events in the defect trajectories, and (consequently) they approximate an effective boundary for the region of observed defect motions. These control mechanisms hold both for the time-periodic ($n\leq 4$) systems and the \removed{non-}\added{a}periodic ($n \geq 5)$ systems; hence, we can use them to understand the defect braiding dynamics that they permit or prohibit.

\subsection*{Flow-field structure}

Nematohydrodynamics produces a two-way coupling between the \removed{structure of the} flow velocity field $\bf u$ and the director field $\bf n$, and we see this reflected in the coupled topologies of the two fields.  In addition to determining the reorientation sites of defect trajectories, the active force encoded by the boundary conditions acts on the flow field. \removed{Thus, understanding how the structure of the flow field is mutually-consistent with the boundary-imposed active force will be useful for characterizing the general dynamics of $n=2q$ positive defects with anchoring of winding $q$.}
Here\added{, we examine the topologically robust features of the active nematic flow field that are sculpted by the choice of boundary conditions.}\removed{, by fixing the director at the boundary with nontrivial winding, we set up patterns of active force by which we can sculpt the  dominant flow structures within the bulk.}

To elucidate the coupling of the director and velocity fields, we examine the instantaneous and time-averaged vorticity of the flow structures underlying the observed braiding dynamics. We  define the boundaries of vortices using the second invariant of the velocity gradient tensor, known as the $\mathcal{Q}$-criterion \cite{head_spontaneous_2024}:
\begin{equation}
    \mathcal{Q} = \tfrac{1}{2} \left(||\omega||^2 - ||E||^2\right)
\end{equation}
\added{where $\omega$ and $E$, the vorticity and strain rate tensors, are respectively the antisymmetric and symmetric components of the velocity gradient tensor. }Positive values of $\mathcal{Q}$ correspond to vorticity-dominated regions and negative ones to strain-dominated regions. Thus, the closed isolines of $\mathcal{Q}=0$ provide \added{a suitable and unambiguous definition of}\removed{well-defined} vortex boundaries.  

\begin{figure*}[h!]
    \centering
    \includegraphics[width=\textwidth]{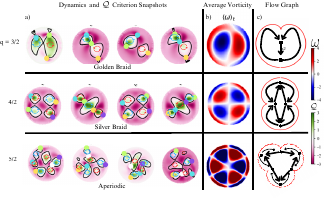}    \caption{
    Coupled topologies of flow fields and topological defect trajectories in active nematics confined to disks with excess topological charge due to anchoring winding $q$ as labeled at left.  (a) Snapshots from time series showing representative dynamics of $+1/2$ defects and flow fields. The magenta-green heatmaps show the $\mathcal{Q}$-criterion of the flow field. Black curves are the viscometric isolines where $\mathcal{Q} = 0$, each enclosing a vortex, whose vorticity \added{$\omega$ }is indicated by blue or red curved arrows. Colored \added{filled }circles mark the instantaneous positions of $+1/2$ defects. \removed{The braids exhibited by the $+1/2$ defects are the golden braid for $q=3/2$ system and the silver braid for $q=4/2$. For $q=5/2$ (and higher) no periodic braid is observed.} (b) The time-averaged vorticity field, showing \removed{2, 4, and 6 gyres for the $q=3/2$, $4/2$, and $5/2$ systems respectively. This counting excludes the change in sign of vorticity in a narrow band near the boundary}\added{flow gyre structure}. (c) Directed graphs schematically summarizing the flow structure mandated by the boundary conditions, containing $4|q-1|$ gyres among its cycles. \removed{Each node has a total degree of 0 as is consistent with incompressible flow.}}
    \label{fig:6}
\end{figure*}

Defect motions and vortex structure are strongly linked through the $\mathcal{Q}$-criterion: the trajectories of $+1/2$ defects in simulated and experimental active nematics data closely follow isolines of $\mathcal{Q} = 0$, an observation rationalized by the fact that  the Stokes-flow solution for an isolated $+1/2$ defect lies on a $\mathcal{Q} = 0$ isoline \cite{head_spontaneous_2024}. We observe this ``self-constraint'' between the $\mathbf{n}$ and $\mathbf{u}$ fields to be well-obeyed in our system, for both periodic and aperiodic motion as exemplified by the snapshots in \removed{Figure}\added{Fig.} \ref{fig:6}a.  The instantaneous and running time-averaged fields of both $\omega$ and $\mathcal{Q}$, along with their standard deviations, are shown for the $q = 3/2$, $4/2$, and $5/2$ systems in Movies S4, S5, and S6 \added{respectively}. 

Importantly, we find that \added{this self-constraint requires that every defect swap in braiding motions coincide with} \removed{necessitates} a topological change of $\mathcal{Q} =0$ isolines: Because each $+1/2$ defect is always at a location with $\mathcal{Q}=0$ instantaneously, and the $\mathcal{Q}=0$ subset consists of closed curves each surrounding a vortex $(\mathcal{Q}>0)$, a defect can only move from one vortex to another by instantaneous intersection of two $\mathcal{Q}=0$ loops into a ``figure eight''. These intersections are visible in some of the snapshots of Fig.~\ref{fig:6}a. In order for the circulation direction of the vortex to be consistent with the $+1/2$ defect's heading, these direct vortex-swapping events only take place between vortices of opposite-sign vorticity. 

However, defects can also swap with one another between same-sign vortices by a more complicated choreography, seen in the $q=4/2$ system (Fig.~\ref{fig:6}a middle row): Two CW vortices merge into one, while coming temporarily into contact with the two CCW vortices. Two defects, previously on the two separate CW vortices, are now pinned to these two $\mathcal{Q}=0$ junctions while continuing to circulate clockwise. Then, the CW vortex splits again into two while breaking contact with the CCW vortices. The flow field has now returned to its original structure, but with two defects having swapped CCW vortices. The mirror-image process then occurs to allow swapping of the other two defects between the two CCW vortices. These alternating swaps produce the two distinct, overlapping defect trajectories in Fig.~\ref{fig:5}c. 

The time-averaged vorticity for these three systems, plotted in Fig.~\ref{fig:6}b, is markedly dominated by $4|q-1|$ alternating gyres of opposite vorticity. The time-averaged behavior dominates the instantaneous dynamics, in the sense that we measure a small noise-to-signal ratio, defined as the spatially averaged ratio between the standard deviation and the time average of the vorticity over all lattice points. These ratios are $0.1048$ for the $3/2$ system, $0.0069$ for the $4/2$ system and $0.0007$ for the $5/2$ system.

 Gyre boundaries divide adjacent regions of alternating time-averaged vorticity, partitioning the domain. Each gyre boundary advects material either toward or away from the fluid boundary layer, where the vorticity changes sign to accommodate the no-slip boundary condition  \cite{shankar2024design, giomi_geometry_2015}. (For $q=3/2$, the single gyre boundary moves material away from one side of the circle and toward the other.)  The locations where gyre boundaries advect material away from the boundary coincide with the angular positions of the cusps in the envelope curves plotted in Fig.~\ref{fig:5}b. Equivalently, \removed{this means that} the lobes between cusps on the envelope \removed{also} each contain two counter-rotating domains but in such a way that the flow advects material along the boundary. This matches our findings that splay deformation fixed by the boundary condition scatters defects into the bulk while fixed bend polarization allows them to travel along the boundary. The cusps thus behave similarly to ``wall defects'' observed in experiment to arise spontaneously in active nematics confined to a disk \cite{hardouin_active_2022}. 

The global consistency of these dynamics can be understood  schematically as a directed graph over each envelope, as shown in Fig.~\ref{fig:6}c. Each node corresponds to either a cusp or a bulge, and directed edges signify the local flow direction\added{;} \removed{such that} the gyre structure is summarized by cycles in the graph. \removed{Importantly, t}\added{T}he net degree of every node is 0, representing the incompressibility constraint. This yields a unique gyre structure for every $q$, in good agreement with that measured in Fig.~\ref{fig:6}b.

 To find a general expression for the number of gyres, we can count the sectors of the active force field at the boundaries as the number of times the active force vector is aligned parallel or antiparallel to $\hat r$ as we vary over orientations in the plane. Thus, the number of gyres is given by the number of solutions to
\removed{\begin{equation}
    \begin{aligned}
           \mathbf r \times \mathbf F_{\mathrm{active}} = S\zeta q \sin(2(q-1)\theta) = 0,
    \end{aligned}
\end{equation}}
\added{$\mathbf r \times \mathbf F_{\mathrm{active}} = S\zeta q \sin(2(q-1)\theta) = 0,$ }from which we find that the number of gyres grows as $4|q-1|$, as we observe numerically (Fig.~\ref{fig:6}b). In contrast, the number of defects grows as $2q$. This leads to the simple but important observation that for $q > 4/2$, there are more gyres than defects. While there may be time-variation in the number of vortices, the active force field required by the boundary conditions demands that the number and structure of vortices give, on average, the gyre structures of Fig.~\ref{fig:6}b. As seen in Fig.~\ref{fig:6}a and Movie S6, vortices without defects are unstable and tend to decay in size; some disappear entirely, to be replaced by nucleation of a new $\mathcal Q = 0$ loop elsewhere in the system. This instability is consistent with the previously observed enhancement of vortex stability by the presence of a $+1/2$ defect \cite{head_spontaneous_2024, giomi_geometry_2015}. 

The presence or absence of such defect-free vortex ``holes'' is important to defect braiding because $+1/2$ defects tend to distribute themselves evenly among the available $\mathcal Q=0$ loops (presumably due, at least in part, to elastic repulsion between like-sign defects) so a hole will be available for a defect to transfer to only if the vortices outnumber the defects. Otherwise, a defect can only transfer to another $\mathcal Q=0$ loop by a coordinated swap with another defect. Sequences of these pairwise swaps are required for periodic braiding and maximal topological entropy production, which thus occurs only for $q=3/2$ and $4/2$. 

Understanding of this complex interplay of flow-field and director topologies is facilitated by \added{interpreting the schematic directed graph of Fig.~\ref{fig:6}c as applying not onlyl to the flow field but also to the defects.} \removed{a schematic representation of defect paths consistent with the time-averaged vorticity. This representation, shown in Fig.~\ref{fig:6}c, takes the form of a directed graph whose edges imply a ``flow'' consistent with the corresponding gyres of Fig.~\ref{fig:6}b. One node is placed for each pair of adjacent, counter-rotating gyres.} A defect encountering a node can exit it along either of two outward-directed edges, one of which keeps the defect on the same gyre, while the other represents a transfer to the neighboring gyre. For $q=3/2$ or $4/2$, such a transfer necessarily causes a temporary ``overcrowding'' of one gyre by two defects, leading to the transfer of the gyre's original defect to another gyre. For $q=3/2$, there is only one other gyre available, while for $q=4/2$, transfer to the single empty gyre is favored, ensuring a pairwise swap of the two transferring defects. \added{For $q=5/2$ (and higher) each defect has multiple empty gyres into which it can jump, so motion through the graph is under-constrained.}

\subsection*{Geometrically structured confinement with tangential anchoring}
\added{Having studied the hypothetical steady-winding circle boundary conditions so far, we now demonstrate the applicability of our findings to closed confinement geometries with  more experimentally realistic boundary conditions of tangential anchoring \cite{doostmohammadi_active_2018} on a geometrically structured boundary. We begin with the cardioid confinement geometry studied experimentally in Ref.~\cite{mitchell_cardioid}, where the cardioid's cusp pins a $-1/2$ so the interior has net topological charge $q=3/2$, requiring an extra $+1/2$ defect besides the two already present for circular confinement. We also examine generalizations of this geometry to two and three evenly spaced cusps using the epicycloid curves introduced above as boundaries. Since each cusp pins a $-1/2$ defect, every additional cusp adds $+1/2$ to the net topological charge of the interior, allowing control over the number or $+1/2$ defects.}

\begin{figure*}[h!]
    \centering
    \includegraphics[width=\textwidth]{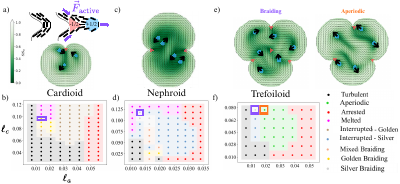}
    \caption{\added{Simulated active Beris-Edwards nematohydrodynamics in domains with  tangential-anchoring boundaries containing one or more regularized cusps: the cardioid (a,b), nephroid (c, d), and trefoiloid (e, f).} a, c\added{, e}) Snapshots from \added{representative cases of the }braiding\added{ and aperiodic} regimes\removed{.}\added{, at } $(\ell_a,\ell_c) $ values \added{indicated by purple or orange boxes in (b), (d), and (f) respectively:} \removed{are} $(0.0139, 0.0903)$ for the cardioid\added{,} \removed{(left) and} $(0.0131, 0.1178)$ for the nephroid\added{,} \removed{(right}\added{and $(0.0128, 0.0766)$ and $(0.0192, 0.0766)$ respectively for the braiding and aperiodic cases shown for the trefoiloid.} \removed{showing pinned negative defects at epitrochoid cusps and motile defects in respective golden and silver braiding patterns.} \added{All reported braids were obtained consistently over 5-10 random initializations per pair of $(\ell_a, \ell_c)$ values.} \removed{Bottom}\added{Top} inset of a)  schematically illustrates defect pinning on concave boundary features. The outwards facing normal vectors are shown as black arrows. The left image shows a non-pinning director configuration, in which there is a \removed{single} bend distortion from the boundary. The right image shows \added{a $-1/2$ defect pinned at the boundary and a nearby} \removed{defect pinning whereby the bend wall shown on the left nucleates a defect pair and pins a $-1/2$ defect against the concave boundary, and the created} $+1/2$ defect \removed{moves}\added{moving} into the nematic bulk. b, d\added{, f}) Active phase diagrams for the cardioid (\removed{left}\added{b}) \removed{and}\added{,} nephroid (\removed{right}\added{d), and trefoiloid (f)} epi\removed{trochoid}\added{cycloid} curves\removed{ using continuum Beris-Edwards nematohydrodynamics}. \removed{The stability of the braiding patterns are highly dependent on the confining geometries.} Cardioid\added{ and trefoiloid} simulations \added{were }performed on a $200\times 200$ lattice and nephroid simulations \added{were }performed on a $100\times 100$ lattice\added{,} each for $1.5\times 10^6$ time-steps. \removed{Simulations}\added{States} are labeled turbulent if there is spontaneous pair production in the nematic bulk, arrested if the defects do not move, and melted if the defect cores can not be uniquely distinguished. A braid is labeled \removed{metastable}\added{interrupted} if \added{the pinned $-1/2$ defects briefly absorb and re-emit each $+1/2$ defect during the periodic cycle but the defect trajectories are otherwise topologically identical to the golden or silver braids.} \removed{the dynamics maintain the flow structure of the braids, but are non-pinning and/or non-cyclic, temporarily absorbing and emitting a positive defect at the cusp during the braid cycle through pair-annihilation and subsequent pair-creation of $\pm 1/2 $ defects.} Representative dynamics are shown in Movies S7\added{, S8,} and S\removed{8}\added{9}.}
    \label{fig:7}
\end{figure*}


\added{For the $q=3/2$ epicycloid with one  cusp, we obtain for some parameter values a stable golden braid of three $+1/2$ defects, in agreement with the $q=3/2$ steady-winding circle boundary (Fig.~\ref{fig:7}a). As expected, a $-1/2$ defect is pinned to the cusp at all times.} \removed{Here, this occurs through the pinning of a $-1/2$ defect to each of the (regularized) cusps as seen in \added{Figures} \ref{fig:7}a and \ref{fig:7}c, leaving three or four $+1/2$ defects to interact in the bulk.} The geometry of the $-1/2$ defect reorients the bulk director field nearby, creating the splay mode \removed{seen}\added{just as we saw} at the envelope\added{'s} cusp in \added{steady-winding} circular confinement, \removed{and thus} providing the inward active force that promotes the double \removed{or quadruple} gyre flow structure. 

\added{Interestingly, this system exhibits a stable interrupted regime (Fig.~\ref{fig:7}b), shown in Movie S7, in which each the periodic motion involves each $+1/2$ defect sequentially being pair-annihilated with the cusp's $-1/2$ defect and then effectively re-emitted by defect pair creation at the same location. Within the braiding taxonomy this is simply rectified by permitting that the newly created defect carries on the same braid strands as the recently annihilated defect.}

\added{Similar dynamics are seen in the $q=4/2$ epicycloid (Fig.~\ref{fig:7}c, Movie S8). The silver braid predicted by the steady-winding circle boundary spontaneously emerges as a stable state for a large portion of parameter space. For most of these parameters, the silver braid is interrupted (Fig.~\ref{fig:7}d), with all four $+1/2$ defects being continually absorbed and emitted by the cusps. The splay distortion created by the two cusp-bound $-1/2$ defects provides two sources of strong, inward active force with opposite directions, setting up the double     There are some parameters at which we instead observe a $-1/2$ defect at only one of the two cusps, meaning the interior has a net charge of $+3/2$. In these cases, the three defects follow a golden braid, further demonstrating that this braid is a robust property of three $+1/2$ defects in different boundary geometries. We have observed cases of silver braid transforming after a few cycles into a golden braid (lower right panel of Movie S8).}

In general, \removed{we find that }\added{the active force experienced due to the boundary conditions is qualitatively altered by the pinning of negative defects at cusps. The splay introduced near the cusps results in} \removed{the pinning of negative defects changes the active force produced by the new effective boundary and realizes} the same topology of active force structure described in equation ~\ref{eq:f_active_epicycloid} in circular confinement, and thus the same number of gyres. 

\added{The case of $q = 5/2$ is largely consistent with the predictions of the steady-winding circle:  aperiodic motion, similar to motion in Fig. \ref{fig:5}c, fills most of the parameter space region between the turbulent and arrested states (Fig. \ref{fig:7}f). There is, however, a narrow range of this parameter space where a periodic state emerges: A fourth $-1/2$ defect is dynamically pinned at the geometric center of the system, making the number of $+1/2$ defects six instead of five (Fig. \ref{fig:7}e). 
The $+1/2$ defects circulate in a periodic motion topologically equivalent to the ``Ceilidh dance''~\cite{shendruk_dancing_2017}. In the braiding taxonomy, this is the six-defect case of the silver braid. 
The emergent periodic motion is consistent with our defect-gyre counting criterion, as the number of $+1/2$ defects now equals the number of gyres, which is twice the number of boundary cusps. The central $-1/2$ defect is presumably stabilized by \removed{sharing the same three-fold rotational symmetry as the boundary conditions; also, the six gyres of alternating handedness agree with the active flow vortex structure around an ideal, isolated $-1/2$ defect \cite{giomi_geometry_2015}}\added{the shared active force structure, consistent the fact that the two solutions of $4|q-1| = 6$ are $q = -1/2$ and $q = 5/2$}.
}%
\removed{
Interestingly, this system exhibits a strongly metastable regime, shown in Movies S7 and S8, in which one or two defects are absorbed into a cusp and then emitted during part of the braid cycle. Within the braiding taxonomy this is simply rectified by permitting that the braid strands can jump from the absorbed defect to the corresponding new positive charge once it nucleates from the cusp into the bulk (Fig. \ref{fig:7}a).
}%
\subsection*{Agent-based active nematic filament simulations}
To assess the generality of \removed{the theory}\added{our findings}, we \added{turn to an alternative simulation method that accounts for}\removed{note} the importance of density fluctuations in \added{microtubule-kinesin }active \added{nematic}\removed{microtubule} experiments\added{,} which are absent in the Beris-Edwards model. We simulate a three-dimensional system of active microtubules confined in a thin layer vertically and in a cardioid geometry laterally, using an agent-based, coarse-grained model of active bead-spring filaments as described in 
the Supporting Information. A small thickness\removed{, $L$,} in the third dimension allows filament cross-over and is crucial for reproducing defect dynamics. Uniquely, the active filaments are coupled to a two-dimensional underlying coarse-grained fluid layer. This fluid layer provides two important properties otherwise missing from the coarse-grained active model: long-range hydrodynamic interactions and a momentum-conserving thermostat. The hydrodynamic interactions permit similar density fluctuations to microtubule experiments. Temperature control was achieved using a pairwise dissipative particle dynamics thermostat and was passed through to the active layer of the simulation via an active-fluid particle interaction moderated by an artificial distance offset.

\begin{figure*}[h!]
    \centering
    \includegraphics[width=\textwidth]{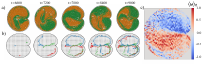}
    \caption{(a) Time series of filament dynamics of the agent-based simulations of confined active nematic filaments inside the cardioid with $4 \times 10^5$ time-steps between each image shown. Green and orange indicate CCW and CW polar orientation with respect to the origin, with underlying fluid particles not visualized. Simulations consists of 800 filaments, each 80 beads long.  Visualization times shown above are in arbitrary simulation time units. (b) Locally-averaged director field corresponding to filament visualizations above. Three long-lived defects were tracked using the director field and their trajectories plotted over the director field showing their path and braiding dynamics. (c) Time-averaged vorticity of agent-based simulations’ filament particle velocities showing double gyre structure with positive vorticity shown in red and negative vorticity shown in blue. Vorticity units are \removed{$\sigma\sqrt{m/\varepsilon}$}\added{$\sigma (m/\varepsilon)^{1/2}$} (see 
    Supporting Information). Full simulation with interpolated director field and extracted defect cores and trajectories is shown in Movie S\removed{9}\added{10}.}
    \label{fig:8}
\end{figure*}

Our agent-based model closely mimics experimental realizations of strongly confined microtubule-kinesin active nematics~\cite{memarian_active_2021,opathalage_self-organized_2019}. Simulations consist of $64,000$ active particles comprising $800$ filaments, each with $80$ beads, and an additional $3068$ fluid particles. The time evolution of these active filaments is shown in \removed{Figure}\added{Fig.}~\ref{fig:8}a and in Movie S9. Upon careful observation of the active filaments, we observe the existence of three, long-lived, $+1/2$  defects as predicted in a system with $q=3/2$. By interpolating bead-spring orientation to a director field, we can more clearly see the $+1/2$ defects and their trajectories in \removed{Figure}\added{Fig.}~\ref{fig:8}b. We track these three defects over a period of simulation time and observe that their motion is consistent with the golden braid as seen in the nematohydrodynamics \added{epicycloid }system with $q=3/2$, in that the defects perform alternating swaps from each half of the cardioid. In addition, we also see the double gyre flow structure from the time-averaged vorticity of the agent-based model (\removed{Figure}\added{Fig.}~\ref{fig:8}c). This double gyre structure also compares favorably to the time-averaged vorticity of the $q=3/2$ nematohydrodynamics model shown in \removed{Figure}\added{Fig.}~\ref{fig:6}b, containing alternating positive and negative values of vorticity in the upper and lower halves of the cardioid. 

Unlike in the continuum nematohydrodynamics model, the agent-based model incorporates non-uniform density, where defects can be annihilated or created on the boundary between empty space and the bulk. As a result, at any given time, there may be an additional defect pair of $+1/2$ and $-1/2$ defects that quickly annihilate. However, low-density regions are a characteristic of the experimental systems of active kinesin-driven microtubules. We predict that the agent-based model, which has an active length scale and nematic coherence length that are difficult to estimate, is close to the boundary between golden braiding and interrupted states in Fig~\ref{fig:7}b.

\section*{Discussion}
\added{We have computationally studied the conditions necessary for time-periodic dynamics to emerge in confined two-dimensional active nematics, and we have identified the specific braiding motions of $+1/2$ defects in these dynamics. Our findings show that the golden braid observed experimentally in cardioid-shaped confinement \cite{mitchell_cardioid} is  generic to closed geometries with exactly three $+1/2$ defects in the interior. Not only did this braid arise in the $q=3/2$ steady-winding circle and our simulations of the cardioid, but we also observed it in the  $q=4/2$ epicycloid geometry when there were three $+1/2$ defects because only one of the two regularized cusps pinned a $-1/2$ defect.}

\added{Likewise, we predict that a closed geometry with exactly four $+1/2$ defects in the interior will exhibit the silver braid, which has not yet been observed in experiments but which we found in the $q=4/2$ steady-winding circle and in the $q=4/2$ epicycloid when both cusps pinned a $-1/2$ defect. We note that an effective quasiparticle model for $+1/2$ defects in active nematics \cite{cody_analytic} found the golden braid for $n=3$ but did not observe the silver braid for $n=4$. It is interesting to note that quasiparticle models for \emph{vortices} have similarly found a cross-over from periodic to unstable dynamics a small number of particles on 2-dimensional manifolds. \cite{aref2007point, aref1988point}}

\added{Other behaviors, namely   active turbulence, arrested states, and melted states, were observed in the expected parameter regimes. Importantly, however, the golden and silver braids were the only motions found when there were respectively $n=3$ or $4$ motile positive defects, and no negative defects, in the interior.}

\added{For $q=5/2$ and higher, the situation is more subtle but ultimately illuminates the underlying mechanism. Only aperiodic dynamics are observed in the steady-winding circle system. Aperiodic dynamics also dominate in the $q=5/2$ epicycloid, but in a narrow region of parameter space we find a new periodic state, a silver braid of six $+1/2$ defects dancing around an extra $-1/2$ defect in the center (Fig. \ref{fig:7}e).}

\added{We rationalized these observations by establishing topological connections between the boundary conditions on the director field and the gyre structure produced in the time-averaged velocity field by the active forces imposed by the boundary condition. We noted that a net topological charge $q$ in the interior also requires that there be $2|q-1|$ maxima of splay distortion at the boundary where the active force is strongly inward, alternating with an equal number of bend distortion maxima where the active force is strongly outward (Fig.~\ref{fig:5}c). Therefore, the boundary promotes $4|q-1|$ flow gyres of alternating vorticity (Fig.~\ref{fig:6}b,c). For the epicycloids, the $-1/2$ defects pinned at cusps became part of the effective boundary, each introducing an extra pair of gyres. For $q=3/2$, the double gyre is seen in cardioid-shaped confinement in experiment \cite{mitchell_cardioid} and here in two simulation frameworks (Movie S7, Fig.~\ref{fig:8}c), as well as in our simulations of the steady-winding circle system (Figs.~\ref{fig:6}b). }

\added{The connection between the flow gyre structure and defect trajectories is provided by the rule proposed in Ref.~\cite{head_spontaneous_2024} that $+1/2$ defects are constrained to $\mathcal{Q}=0$ isolines which form closed loops at vortex boundaries. Our results are consistent with this rule and reveal that this ``self-constraint'' interestingly requires that defect swaps, the building blocks of the braiding motions, must coincide with topological changes in the flow structure, in the form of reconnections in the $\mathcal{Q}=0$ isolines. \removed{Assuming that the instantaneous vortex structure consists primarily of small deviations from the time-averaged gyre structure, and using the observation that two $+1/2$ defects can share a single vortex only very briefly, we are led to our proposed general rule: Periodic defect motions arise if and only if the $+1/2$ defects are at least as numerous as the flow gyres, as this constrains the possible defect swaps to certain periodic cycles. When there no $-1/2$ defects in the  interior, the number of $+1/2$ defects is $n=2q$  while  the number of gyres is $4|q-1|$. This leads to the prediction of  periodic dynamics up to $n=4$ defects and aperiodic dynamics for $n \geq 5$, in agreement with our simulations when the interior had no $-1/2$ defects. The qualitative agreement of our agent-based model with the defect braiding and flow gyre structure predicted by our theory demonstrates the robustness of this topological approach. }}

\added{
Our counting rule  of $4|q-1|$ gyres is consistent with previous findings for active nematics in channels with periodic boundary conditions, where the stability of the ``Ceilidh dance'' is explainable by its equal numbers of $+1/2$ defects and gyres \cite{shendruk_dancing_2017}. We can interpret that dance as \removed{equivalent to }a silver braid, as we observed in circular confinement with $q=4/2$ winding boundary conditions, by tiling the channel with interconnected unit cells of the braiding domain. Further, \added{our results provide a systematic connection between boundary geometry and stable vortex structures via the active force density in a variety of active nematic systems, including} \removed{the presence of coherent flow structures  in confined active nematics has been seen in} channel confinement \cite{Wagner_coherent, hardouin_active_2022} \removed{as well as under}\added{and} spatially periodic patterning of activity and external forces \cite{calderer_chevron_2024, Mozaffari_spirograph}. \added{It would be useful in future work to consider deformable or noisy boundary conditions \cite{ben2022disordered, velez2024probing} to characterize the full extent of the topological connection between average director and flow structures.}
}

\added{
\removed{Additional control mechanisms such as internal obstacles or inclusions could potentially aid in this stability.}
Our findings could help to guide extensions to recent advances in this regard, such as the use of spatially periodic tilings of obstacles to create active pumps \cite{velez2024active} or vortex lattices with effective anti-ferromagnetic ordering \cite{schimming2024vortex}.
 \removed{Ordered flow structures in active nematics can in turn set up ordered flows in passive liquid crystals with which they are in contact \cite{calderer_chevron_2024}, which may be used to guide micro-swimmers in living liquid crystals \cite{turiv2020polar}.} It is also possible that inclusions which contain fixed interior winding could be used to generate larger arrays of isolated periodic orbits connected through a nematic bulk, which could be useful in efforts to realize increasingly complicated time-crystal behaviors in soft matter \cite{zhao2025space, Morrell2026}.
}

We have demonstrated numerically that the spontaneous golden braiding of defects in cardioid confinement \removed{(Ref.}~\cite{mitchell_cardioid}\removed{)} are one example of a more general phenomenon, in which active nematics spontaneously adopt \removed{optimally mixing defect}\added{periodic} braiding motions \added{of defects }if these are consistent with, and suitably constrained by, the topology of vortices in the flow field. \added{Among the possible periodic braids, optimal stirring braids seem to be preferred for small numbers of defects.} \removed{In confinement with circular geometry and controllable winding in the anchoring, we observed these maximally mixing periodic states in the form of the golden braid for $n=3$ defects and the silver braid for $n=4$.}\removed{
 Not only are these braids consistent with those we find with tangential-anchoring boundaries with $n-2$ cusps, but we obtained an emergent rationale for this connection through the resemblance of the latter boundaries to the envelope curves of the variable anchoring in our circular-boundary systems. Those envelopes govern defect trajectories for both periodic ($n\leq 4$) and aperiodic ($n\geq 5$) defect motions. The effective boundaries created by the envelopes, together with the gyre structures required by active forces at the boundaries, provide new principles for geometrically and topologically controlling the motions of topological defects }
 \added{Our proposed criterion for periodic motion, based on the motile defects being at least as numerous as the flow gyres, offers a systematic principle to guide the design of ordered flows} in active nematics, potentially opening new avenues for fluid mixing applications.



\subsubsection*{SI Movies}

\def\matmethods{
\subsection*{Beris-Edwards nematohydrodynamics}
We computationally model an extensile active nematic in two dimensions, with nematic order represented by the second-rank tensor
\begin{equation}
Q_{ij}(\mathbf{r}) = S(\mathbf{r})\left(n_i(\mathbf{r})  n_j (\mathbf{r})- \tfrac{1}{2}\delta_{ij}\right),
\label{eq:Q_def}
\end{equation}
where $\bf n$ is the director, $\delta_{ij}$ is the Kronecker delta, $i$ and $j$ run over two dimensions, and $S$ is the scalar degree of nematic order.

We simulate Beris-Edwards nematohydrodynamics \cite{beris1994thermodynamics} in a finite difference scheme on a square lattice, similarly to the approach of Ref.~\cite{giomi_geometry_2015}, describing the coupled evolution of the nematic order, $Q_{ij}$, and the flow field, $\bf u$. The time-evolution of $Q_{ij}$ is governed by
\begin{equation}
    \begin{aligned}
        \partial_t Q_{ij} + u_k\partial_k Q_{ij}=  \frac{1}{\gamma}H_{ij} + \chi S E_{ij} + [Q, \omega]_{ij} - 2\mathrm{Tr}[QE]Q_{ij}. \label{eq: dtQ}
    \end{aligned}
\end{equation}
Here, $[\cdot,\cdot]$ is the commutator, $\bf u$ is the flow field, $\gamma$ is the rotational viscosity, $\chi$ is a flow aligning or tumbling parameter, and $E_{ij} = (\partial_i u_j + \partial_j u_i)/2$ and $\omega_{ij} = (\partial_i u_j - \partial_j u_i)/2$ are the rate-of-strain and vorticity tensors respectively. $H_{ij}$ is the molecular tensor associated with the Landau-de Gennes free energy 
\removed{\begin{equation}
    \begin{aligned}
    F_{\mathrm{LdG}} = \int_\Omega \biggl[ \frac{A}{2}\mathrm{Tr}[Q^2]\left(1 - \frac{1}{2}\mathrm{Tr}[Q^2]\right) 
    \\+ \frac{K}{2}(\nabla_i Q_{jk})(\nabla_i Q_{jk})  \biggr] \; d\Omega,
\end{aligned}
\end{equation}}
\added{
\begin{align}
 F_{\mathrm{LdG}} &= \tfrac{1}{2} K \int_\Omega \biggl[ \ell_c^{-2} \mathrm{Tr}[Q^2]\left(1 - \tfrac{1}{2}\mathrm{Tr}[Q^2]\right) 
    + (\nabla_i Q_{jk})(\nabla_i Q_{jk})  \biggr] \; d\Omega,
\nonumber \\
H_{ij} &= - \frac{\delta F_{\mathrm{LdG}}}{\delta Q_{ij}} = -Q_{ij} A(1 - \mathrm{Tr}[Q^2]) + K \nabla^2 Q_{ij}.
\end{align}
}
\removed{\begin{equation}
    \begin{aligned}
H_{ij} &= - \frac{\delta F_{\mathrm{LdG}}}{\delta Q_{ij}} \\ &= -Q_{ij} A(1 - \mathrm{Tr}[Q^2]) + K \nabla^2 Q_{ij}.
\end{aligned}
\end{equation}}
Here, \removed{$A < 0$ and sets the energy cost of a defect,}\added{the nematic correlation length $\ell_c$ sets the typical size of a defect core,} $K$ is the Frank elastic constant in the one-constant approximation, and $\Omega$ is the nematic domain.

The time evolution of $\bf u$ is given by the incompressible Navier-Stokes equations, 
\removed{
\begin{align}
    \partial_t {\bf u} +({\bf u} \cdot \nabla){\bf u} &=   \eta \nabla^2 {\bf u} + \frac{1}{\rho}{\bf F} - \frac{1}{\rho}\nabla { p}, \\
\nabla \cdot {\bf u} &= 0
\end{align}
}
\added{\begin{align}
    \partial_t {\bf u} +({\bf u} \cdot \nabla){\bf u} &=   \eta \nabla^2 {\bf u} + \frac{1}{\rho}{\bf F} - \frac{1}{\rho}\nabla { p}, \label{eq: navier-stokes} \qquad 
\nabla \cdot {\bf u} = 0,
\end{align}}
where $\eta$ is the fluid viscosity, $\rho$ is the (constant) density, and the force density $\bf F$ is the divergence of a stress tensor $\Pi_{ij}$:
\begin{equation}
    \begin{aligned}
        F_i = \partial_j  \Pi_{ij} = \partial_j [- H_{ij} - \zeta Q_{ij} + [Q, H]_{ij} \\+ 2\mathrm{Tr}[QH]Q_{ij} - K \partial_i Q_{kl} \partial_j Q_{kl}].
    \end{aligned}
\end{equation}
Here, $\zeta$ is the activity, coupling the flow field time evolution to nematic \removed{stresses}\added{distortions}. For all systems considered here, $\zeta >  0$, which corresponds to extensile activity.

This scheme, when simulated in periodic boundaries, reproduces the well-known bulk behavior of ``topological chaos''  \cite{ngo2014large, decamp2015orientational, giomi_geometry_2015, tan_topological_2019}, generating $\pm 1/2$ topological defects corresponding to locally melted regions of the nematic order which couple to the surrounding flow field. \removed{There are two characteristic length scales that emerge from these equations of motion: the active length scale $\ell_a = \sqrt{K/\zeta}$ which scales with the average defect spacing, and the nematic coherence length $\ell_c = \sqrt{K/|A|}$ which scales with the average defect core radius.} For each studied geometry, we tune the activity within a range such that $+1/2$ defects are motile and fixed in number, with total topological charge $q$ determined by the geometry. The activity is kept low enough to prevent spontaneous defect pair production, as the regime of active turbulence is not the focus of this study. We use the following values throughout: $\rho = 1, \chi = 1$, $K = 2^{14}$, $\gamma = 100$, and $\eta =\sqrt{10\cdot2^{14}}$.

For simplicity, we take anchoring at the boundaries to be infinitely strong, creating Dirichlet boundary conditions on $Q$. For the velocity field, a no-slip condition $\mathbf{u}=\mathbf{0}$ is applied at the boundaries. 

More details are given in the Supporting Information.

\subsection*{Fluid \removed{mixing}\added{stirring} and defect braiding\label{sec: methods-braiding}}
\removed{
We can describe the exponential stretching produced by defect braiding by considering an infinitely elastic boundary enclosing the defects, which stretches as the defects or stirring rods stretch the initial boundary over time. A lower bound on this stretching, corresponding to pulling the boundary line taught around the defect cores, is given by an exponential function of time from an initial length $L_0$:
\begin{equation}
L(t) = L_0e^{ht}.
\end{equation}
Here, the exponential rate of stretching $h$ is known as the \emph{topological entropy} (Figure \ref{fig:1}a). This scheme for viewing topological defect dynamics is useful for several reasons. Firstly, the topological entropy is an upper bound on the Lyapunov exponent \cite{thiffeault_topology_2006} (Figure \ref{fig:1}b), and in experiments  on microtubule-based active nematics, this bound is nearly met \added{\cite{tan_topological_2019}}. Secondly, it is possible to analytically calculate topological entropy in periodic flows. Hence, we can use $h$ as a proxy for $\lambda$, and measure the chaotic character of active nematics by measuring the stretching injected into the bulk fluid.}

Analytical calculation of \added{the topological entropy }$h$ relies on the description of worldlines of $n$ positive defects within the Artin \removed{B}\added{b}raid group ${\bf B}_n$ \cite{thiffeault_measuring_2005}. The relationship between braiding of defect worldlines and fluid mixing that we calculate assumes that the positive defects act as stirring rods, meaning that their velocity is equal to the local fluid velocity as would be the case if they were solid objects \cite{tan_topological_2019}. \added{Details are given in the Supporting Information.}

 We numerically compute the topological entropy for the two-dimensional active flow by two schemes using the flow field. In the first method, known as the \textit{Line Stretching} algorithm, we advect an initial line segment forward in time according to the local flow, and track the length of the newly advected contour. If the advected segment grows exponentially in time, then the slope from the semi-log plots of contour length over time yields the topological entropy (Movies S2 and S3). Secondly, we use a computational geometry-based algorithm, known as the \textit{E-tec} (Ensemble-based topological entropy calculation) method \cite{roberts2019ensemble}. We advect an ensemble of randomly initialized passive tracers forward in time. The E-tec algorithm computes a lower bound on the entire system’s topological entropy using the finite trajectories of the random ensemble. Full details are provided in Ref.~\cite{roberts2019ensemble}. To compute the Lyapunov exponent, we randomly choose a pair of passive tracers with a very small initial separation distance and track how their separation distance evolves in time. This method quantifies the largest Lyapunov exponent as it only measures the maximum stretching of two nearby tracers without any restriction on the direction of the stretching. 
}

\subsection*{Acknowledgments}

\def\acknow{B.K.\ and D.A.B.\ thank Louise C.~Head and Cody D.~Schimming for many illuminating conversations about flow structure throughout the development of this work. A.J.S.F.\ thanks Jimmy Gonzalez Nu\~nez for his assistance in running Beris-Edwards nematohydrodynamic simulations. We thank Spencer Smith for access to the E-tec software.
This material is based upon work supported by the National Science Foundation under Grant No.\ DMR-2225543 and the U.S.\ Department of Energy, Office of Science, Office of Basic Energy Sciences program under Award No.\ DE-SC0025803. Part of this research was conducted using Pinnacles (NSF MRI, \# 2019144) at the Cyberinfrastructure and Research Technologies (CIRT) at the University of California, Merced.
}

\acknow

\subsection*{Author contributions}

\authorcontributions

\section*{Materials and Methods}

\matmethods


\bibliography{Bib}

@article{Morrell2026,
  title = {Nonreciprocal Wave-Mediated Interactions Power a Classical Time Crystal},
  volume = {136},
  ISSN = {1079-7114},
  url = {http://dx.doi.org/10.1103/zjzk-t81n},
  DOI = {10.1103/zjzk-t81n},
  number = {5},
  journal = {Physical Review Letters},
  publisher = {American Physical Society (APS)},
  author = {Morrell,  Mia C. and Elliott,  Leela and Grier,  David G.},
  year = {2026},  
}

@article{Ramaswamy2017,
  title = {Active matter},
  volume = {2017},
  ISSN = {1742-5468},
  url = {http://dx.doi.org/10.1088/1742-5468/aa6bc5},
  DOI = {10.1088/1742-5468/aa6bc5},
  number = {5},
  journal = {Journal of Statistical Mechanics: Theory and Experiment},
  publisher = {IOP Publishing},
  author = {Ramaswamy,  Sriram},
  year = {2017},
  pages = {054002}
}

@article{aref1988point,
  title={Point vortex dynamics: recent results and open problems},
  author={Aref, Hassan and Kadtke, James B and Zawadzki, Ireneusz and Campbell, Laurence J and Eckhardt, Bruno},
  journal={Fluid Dynamics Research},
  volume={3},
  number={1-4},
  pages={63--74},
  year={1988},
  url={https://doi.org/10.1016/0169-5983(88)90044-5},
  doi={10.1016/0169-5983(88)90044-5}
}

@article{aref2007point,
  title={Point vortex dynamics: a classical mathematics playground},
  author={Aref, Hassan},
  journal={Journal of mathematical Physics},
  volume={48},
  number={6},
  year={2007},
  publisher={AIP Publishing},
  doi={10.1063/1.2425103},
  url={https://doi.org/10.1063/1.2425103}
}

@article{keber2014topology,
  title={Topology and dynamics of active nematic vesicles},
  author={Keber, Felix C and Loiseau, Etienne and Sanchez, Tim and DeCamp, Stephen J and Giomi, Luca and Bowick, Mark J and Marchetti, M Cristina and Dogic, Zvonimir and Bausch, Andreas R},
  journal={Science},
  volume={345},
  number={6201},
  pages={1135--1139},
  year={2014},
  publisher={American Association for the Advancement of Science}
}

@article{marenduzzo2007steady,
	author = {Marenduzzo, D and Orlandini, E and Cates, ME and Yeomans, JM},
	journal = {Physical Review E},
	number = {3},
	pages = {031921},
	publisher = {APS},
	title = {Steady-state hydrodynamic instabilities of active liquid crystals: Hybrid lattice Boltzmann simulations},
	volume = {76},
	year = {2007},
}

@article{duclos2020topological,
	author = {Duclos, Guillaume and Adkins, Raymond and Banerjee, Debarghya and Peterson, Matthew SE and Varghese, Minu and Kolvin, Itamar and Baskaran, Arvind and Pelcovits, Robert A and Powers, Thomas R and Baskaran, Aparna and Toschi, Federico and Hagan, Michael F and Streichan, Sebastian J and Vitelli, Vincenzo and Beller, Daniel A and Dogic, Zvonimir},
	journal = {Science},
	number = {6482},
	pages = {1120--1124},
	publisher = {American Association for the Advancement of Science},
	title = {Topological structure and dynamics of three-dimensional active nematics},
	url = {https://science.sciencemag.org/content/367/6482/1120},
	volume = {367},
	year = {2020}}

@article{Giomi2014,
  title = {Defect dynamics in active nematics},
  volume = {372},
  ISSN = {1471-2962},
  url = {http://dx.doi.org/10.1098/rsta.2013.0365},
  DOI = {10.1098/rsta.2013.0365},
  number = {2029},
  journal = {Philosophical Transactions of the Royal Society A: Mathematical,  Physical and Engineering Sciences},
  publisher = {The Royal Society},
  author = {Giomi,  Luca and Bowick,  Mark J and Mishra,  Prashant and Sknepnek,  Rastko and Cristina Marchetti,  M},
  year = {2014},
  pages = {20130365}
}

@article{RevModPhys.85.1143,
  title = {Hydrodynamics of soft active matter},
  author = {Marchetti, M. C. and Joanny, J. F. and Ramaswamy, S. and Liverpool, T. B. and Prost, J. and Rao, Madan and Simha, R. Aditi},
  journal = {Rev. Mod. Phys.},
  volume = {85},
  issue = {3},
  pages = {1143--1189},
  numpages = {0},
  year = {2013},
  publisher = {American Physical Society},
  doi = {10.1103/RevModPhys.85.1143},
  url = {https://link.aps.org/doi/10.1103/RevModPhys.85.1143}
}

@article{Bestvina_1992,
	author = {Bestvina, Mladen and Handel, Michael},
	title = {Train Tracks and Automorphisms of Free Groups},
	journal = {The Annals of Mathematics},
	year = 1992,
	volume = 135,
	number = 1,
	pages = 1,
	issn = {0003-486X},
	doi = {10.2307/2946562},
	url = {http://dx.doi.org/10.2307/2946562},
	publisher = {JSTOR}
}

@article{sanchez2012spontaneous,
  title={Spontaneous motion in hierarchically assembled active matter},
  author={Sanchez, Tim and Chen, Daniel TN and DeCamp, Stephen J and Heymann, Michael and Dogic, Zvonimir},
  journal={Nature},
  volume={491},
  number={7424},
  pages={431--434},
  year={2012},
  publisher={Nature Publishing Group UK London},
  url={https://doi.org/10.1038/nature11591},
  doi={10.1038/nature11591}
  
}

@article{
kumar_2018_tunable,
author = {Nitin Kumar  and Rui Zhang  and Juan J. de Pablo  and Margaret L. Gardel },
title = {Tunable structure and dynamics of active liquid crystals},
journal = {Science Advances},
volume = {4},
number = {10},
pages = {eaat7779},
year = {2018},
doi = {10.1126/sciadv.aat7779},
url={https://doi.org/10.1126/sciadv.aat7779}
}

@article{zhao2025space,
  title={Space-time crystals from particle-like topological solitons},
  author={Zhao, Hanqing and Smalyukh, Ivan I},
  journal={Nature Materials},
  volume={24},
  number={11},
  pages={1802--1811},
  year={2025},
  publisher={Nature Publishing Group UK London},
  url={https://doi.org/10.1038/s41467-022-34336-z},
  doi={10.1038/s41467-022-34336-z}
}

@article{giomi_geometry_2015,
	title = {Geometry and {Topology} of {Turbulence} in {Active} {Nematics}},
	volume = {5},
	url = {https://doi.org/10.1103/PhysRevX.5.031003},
	doi = {10.1103/PhysRevX.5.031003},
	number = {3},
	urldate = {2024-06-11},
	journal = {Physical Review X},
	author = {Giomi, Luca},
	month = jul,
	year = {2015},
	pages = {031003},
}

@article{bowick2022symmetry,
  title={Symmetry, thermodynamics, and topology in active matter},
  author={Bowick, Mark J and Fakhri, Nikta and Marchetti, M Cristina and Ramaswamy, Sriram},
  journal={Physical Review X},
  volume={12},
  number={1},
  pages={010501},
  year={2022},
  publisher={APS}
}

@article{finn_topological_2011,
	title = {Topological optimisation of rod-stirring devices},
	volume = {53},
	issn = {0036-1445, 1095-7200},
	url = {https://doi.org/10.1137/100791828},
	doi = {10.1137/100791828},
	
	number = {4},
	urldate = {2024-06-11},
	journal = {SIAM Review},
	author = {Finn, Matthew D. and Thiffeault, Jean-Luc},
	month = jan,
	year = {2011},
	pages = {723--743},
}

@article{head_spontaneous_2024,
	title = {Spontaneous {Self}-{Constraint} in {Active} {Nematic} {Flows}},
	volume = {20},
	issn = {1745-2473, 1745-2481},
	url = {https://doi.org/10.1038/s41567-023-02336-5},
	doi = {10.1038/s41567-023-02336-5},
	
	number = {3},
	urldate = {2024-06-11},
	journal = {Nature Physics},
	author = {Head, Louise C. and Dore, Claire and Keogh, Ryan and Bonn, Lasse and Doostmohammadi, Amin and Thijssen, Kristian and Lopez-Leon, Teresa and Shendruk, Tyler N.},
	month = mar,
	year = {2024},
	pages = {492--500},
}

@article{hirst_liquid_2017,
	title = {Liquid crystals in living tissue},
	volume = {544},
	issn = {0028-0836, 1476-4687},
	url = {https://doi.org/10.1038/544164a},
	doi = {10.1038/544164a},
	
	number = {7649},
	urldate = {2024-06-11},
	journal = {Nature},
	author = {Hirst, Linda S. and Charras, Guillaume},
	month = apr,
	year = {2017},
	pages = {164--165},
}

@article{thiffeault_topology_2006,
  title={Topology, braids and mixing in fluids},
  author={Thiffeault, Jean-Luc and Finn, Matthew D},
  journal={Philosophical Transactions of the Royal Society A: Mathematical, Physical and Engineering Sciences},
  volume={364},
  url={https://doi.org/10.1098/rsta.2006.1899},
  number={1849},
  pages={3251--3266},
  year={2006},
  publisher={The Royal Society London}
}

@article{cui_effects_2017,
	title = {Effects of {Anchoring} {Boundary} {Conditions} on {Active} {Nematics}},
	volume = {13},
	issn = {15734137},
	url = {http://dx.doi.org/10.2174/1573413713666170503111004},
	
	number = {4},
	urldate = {2024-06-11},
	journal = {Current Nanoscience},
	author = {Cui, Zhenlu and Su, Jianbing and Zeng, Xiaoming},
	month = jul,
	year = {2017},
}

@article{doostmohammadi_stabilization_2016,
	title = {Stabilization of active matter by flow-vortex lattices and defect ordering},
	volume = {7},
	issn = {2041-1723},
	url = {https://doi.org/10.1038/ncomms10557},
	doi = {10.1038/ncomms10557},
	
	number = {1},
	urldate = {2024-06-11},
	journal = {Nature Communications},
	author = {Doostmohammadi, Amin and Adamer, Michael F. and Thampi, Sumesh P. and Yeomans, Julia M.},
	month = feb,
	year = {2016},
	pages = {10557},
}

@article{duclos_topological_2017,
	title = {Topological defects in confined populations of spindle-shaped cells},
	volume = {13},
	issn = {1745-2473, 1745-2481},
	url = {https://doi.org/10.1038/nphys3876},
	doi = {10.1038/nphys3876},
	
	number = {1},
	urldate = {2024-06-11},
	journal = {Nature Physics},
	author = {Duclos, Guillaume and Erlen\"amper, Christoph and Joanny, Jean-François and Silberzan, Pascal},
	month = jan,
	year = {2017},
	pages = {58--62},
}

@article{shimaya_tilt-induced_2022,
	title = {Tilt-induced polar order and topological defects in growing bacterial populations},
	volume = {1},
	copyright = {https://creativecommons.org/licenses/by-nc-nd/4.0/},
	issn = {2752-6542},
	url = {https://doi.org/10.1093/pnasnexus/pgac269},
	doi = {10.1093/pnasnexus/pgac269},
	
	number = {5},
	urldate = {2024-06-11},
	journal = {PNAS Nexus},
	author = {Shimaya, Takuro and Takeuchi, Kazumasa A},
	editor = {Fratzl, Peter},
	month = nov,
	year = {2022},
	pages = {pgac269},
}

@article{mitchell_maximally_2024,
	title = {Maximally mixing active nematics},
	volume = {109},
	issn = {2470-0045, 2470-0053},
	url = {https://doi.org/10.1103/PhysRevE.109.014606},
	doi = {10.1103/PhysRevE.109.014606},
	
	number = {1},
	urldate = {2024-06-11},
	journal = {Physical Review E},
	author = {Mitchell, Kevin A. and Sabbir, Md Mainul Hasan and Geumhan, Kevin and Smith, Spencer A. and Klein, Brandon and Beller, Daniel A.},
	month = jan,
	year = {2024},
	pages = {014606},
}

@article{thiffeault_measuring_2005,
	title = {Measuring {Topological} {Chaos}},
	volume = {94},
	copyright = {http://link.aps.org/licenses/aps-default-license},
	issn = {0031-9007, 1079-7114},
	url = {https://doi.org/10.1103/PhysRevLett.94.084502},
	doi = {10.1103/PhysRevLett.94.084502},
	
	number = {8},
	urldate = {2024-06-11},
	journal = {Physical Review Letters},
	author = {Thiffeault, Jean-Luc},
	month = mar,
	year = {2005},
	pages = {084502},
}

@article{hokmabad_topological_2019,
	title = {Topological {Stabilization} and {Dynamics} of {Self}-{Propelling} {Nematic} {Shells}},
	volume = {123},
	issn = {0031-9007, 1079-7114},
	url = {https://doi.org/10.1103/PhysRevLett.123.178003},
	doi = {10.1103/PhysRevLett.123.178003},
	
	number = {17},
	urldate = {2024-06-11},
	journal = {Physical Review Letters},
	author = {Hokmabad, Babak Vajdi and Baldwin, Kyle A. and Kr\"uger, Carsten and Bahr, Christian and Maass, Corinna C.},
	month = oct,
	year = {2019},
	pages = {178003},
}

@article{doostmohammadi_active_2018,
	title = {Active nematics},
	volume = {9},
	issn = {2041-1723},
	url = {https://doi.org/10.1038/s41467-018-05666-8},
	doi = {10.1038/s41467-018-05666-8},
	
	number = {1},
	urldate = {2024-06-11},
	journal = {Nature Communications},
	author = {Doostmohammadi, Amin and Ign\'es-Mullol, Jordi and Yeomans, Julia M. and Sagu\'es, Francesc},
	month = aug,
	year = {2018},
	pages = {3246},
}

@article{hardouin_active_2022,
	title = {Active boundary layers in confined active nematics},
	volume = {13},
	issn = {2041-1723},
	url = {https://doi.org/10.1038/s41467-022-34336-z},
	doi = {10.1038/s41467-022-34336-z},
	
	number = {1},
	urldate = {2024-06-11},
	journal = {Nature Communications},
	author = {Hardo\"uin, Jerôme and Dor\'e, Claire and Laurent, Justine and Lopez-Leon, Teresa and Ign\'es-Mullol, Jordi and Sagu\'es, Francesc},
	month = nov,
	year = {2022},
	pages = {6675},
}

@article{tan_topological_2019,
	title = {Topological chaos in active nematics},
	volume = {15},
	issn = {1745-2473, 1745-2481},
	url = {https://doi.org/10.1038/s41567-019-0600-y},
	doi = {10.1038/s41567-019-0600-y},
	number = {10},
	urldate = {2024-06-11},
	journal = {Nature Physics},
	author = {Tan, Amanda J. and Roberts, Eric and Smith, Spencer A. and Olvera, Ulyses Alvarado and Arteaga, Jorge and Fortini, Sam and Mitchell, Kevin A. and Hirst, Linda S.},
	month = oct,
	year = {2019},
	pages = {1033--1039},
}

@article{copenhagen_topological_2021,
	title = {Topological defects promote layer formation in {Myxococcus} xanthus colonies},
	volume = {17},
	issn = {1745-2473, 1745-2481},
	url = {https://doi.org/10.1038/s41567-020-01056-4},
	doi = {10.1038/s41567-020-01056-4},
	number = {2},
	urldate = {2024-06-11},
	journal = {Nature Physics},
	author = {Copenhagen, Katherine and Alert, Ricard and Wingreen, Ned S. and Shaevitz, Joshua W.},
	month = feb,
	year = {2021},
	pages = {211--215},
}

@article{maroudas-sacks_topological_2021,
	title = {Topological defects in the nematic order of actin fibres as organization centres of {Hydra} morphogenesis},
	volume = {17},
	issn = {1745-2473, 1745-2481},
	url = {https://doi.org/10.1038/s41567-020-01083-1},
	doi = {10.1038/s41567-020-01083-1},
	number = {2},
	urldate = {2024-06-11},
	journal = {Nature Physics},
	author = {Maroudas-Sacks, Yonit and Garion, Liora and Shani-Zerbib, Lital and Livshits, Anton and Braun, Erez and Keren, Kinneret},
	month = feb,
	year = {2021},
	pages = {251--259},
}

@article{serra_defect-mediated_2023,
	title = {Defect-mediated dynamics of coherent structures in active nematics},
	volume = {19},
	issn = {1745-2473, 1745-2481},
	url = {https://doi.org/10.1038/s41567-023-02062-y},
	doi = {10.1038/s41567-023-02062-y},
	
	number = {9},
	urldate = {2024-06-11},
	journal = {Nature Physics},
	author = {Serra, Mattia and Lemma, Linnea and Giomi, Luca and Dogic, Zvonimir and Mahadevan, L.},
	month = sep,
	year = {2023},
	pages = {1355--1361},
}

@article{armengol-collado_epithelia_2023,
	title = {Epithelia are multiscale active liquid crystals},
	volume = {19},
	issn = {1745-2473, 1745-2481},
	url = {https://doi.org/10.1038/s41567-023-02179-0},
	doi = {10.1038/s41567-023-02179-0},
	
	number = {12},
	urldate = {2024-06-11},
	journal = {Nature Physics},
	author = {Armengol-Collado, Josep-Maria and Carenza, Livio Nicola and Eckert, Julia and Krommydas, Dimitrios and Giomi, Luca},
	month = dec,
	year = {2023},
	pages = {1773--1779},
}

@article{hardouin_reconfigurable_2019,
	title = {Reconfigurable flows and defect landscape of confined active nematics},
	volume = {2},
	issn = {2399-3650},
	url = {https://doi.org/10.1038/s42005-019-0221-x},
	doi = {10.1038/s42005-019-0221-x},
	
	number = {1},
	urldate = {2024-06-11},
	journal = {Communications Physics},
	author = {Hardo\"uin, J\'erôme and Hughes, Rian and Doostmohammadi, Amin and Laurent, Justine and Lopez-Leon, Teresa and Yeomans, Julia M. and Ign\'es-Mullol, Jordi and Sagu\'es, Francesc},
	month = oct,
	year = {2019},
	pages = {121},
}

@article{smith_braiding_2022,
	title = {Braiding {Dynamics} in {Active} {Nematics}},
	volume = {10},
	issn = {2296-424X},
	url = {https://doi.org/10.3389/fphy.2022.880198},
	doi = {10.3389/fphy.2022.880198},
	
	urldate = {2024-06-11},
	journal = {Frontiers in Physics},
	author = {Smith, Spencer Ambrose and Gong, Ruozhen},
	month = jun,
	year = {2022},
	pages = {880198},
}

@article{smith_topological_2022,
	title = {Topological {Entropy} of {Surface} {Braids} and {Maximally} {Efficient} {Mixing}},
	volume = {21},
	issn = {1536-0040},
	url = {https://doi.org/10.1137/21M142647X},
	doi = {10.1137/21M142647X},
	number = {2},
	urldate = {2024-06-14},
	journal = {SIAM Journal on Applied Dynamical Systems},
	author = {Smith, Spencer A. and Dunn, Sierra},
	year = {2022},
	pages = {1209--1244},
}

@article{shendruk_dancing_2017,
	title = {Dancing disclinations in confined active nematics},
	volume = {13},
	issn = {1744-6848},
	url = {https://doi.org/10.1039/C6SM02310J},
	doi = {10.1039/C6SM02310J},
	number = {21},
	urldate = {2024-06-19},
	journal = {Soft Matter},
	author = {Shendruk, Tyler N. and Doostmohammadi, Amin and Thijssen, Kristian and Yeomans, Julia M.},
	year = {2017},
	pages = {3853--3862},
}

@article{opathalage_self-organized_2019,
	title = {Self-organized dynamics and the transition to turbulence of confined active nematics},
	volume = {116},
	url = {https://doi.org/10.1073/pnas.1816733116},
	doi = {10.1073/pnas.1816733116},
	number = {11},
	urldate = {2024-06-19},
	journal = {Proceedings of the National Academy of Sciences},
	author = {Opathalage, Achini and Norton, Michael M. and Juniper, Michael P. N. and Langeslay, Blake and Aghvami, S. Ali and Fraden, Seth and Dogic, Zvonimir},
	year = {2019},
	pages = {4788--4797},
}

@article{mitchell_cardioid,
  title = {Controlling Chaos: Periodic Defect Braiding in Active Nematics Confined to a Cardioid},
  author = {Memarian, Fereshteh L. and Hammar, Derek and Sabbir, Md Mainul Hasan and Elias, Mark and Mitchell, Kevin A. and Hirst, Linda S.},
  journal = {Physical Review Letters},
  volume = {132},
  issue = {22},
  pages = {228301},
  numpages = {6},
  year = {2024},
  publisher = {American Physical Society},
  doi = {10.1103/PhysRevLett.132.228301},
  url = {https://doi.org/10.1103/PhysRevLett.132.228301}
}

@article{cody_analytic,
  title={Analytical model for the motion and interaction of two-dimensional active nematic defects},
  author={Schimming, Cody D and Reichhardt, CJO and Reichhardt, Charles},
  journal={Soft Matter},
  volume={21},
  url={https://doi.org/10.1039/D4SM00956H},
  number={1},
  pages={122--136},
  year={2025},
  publisher={Royal Society of Chemistry}
}

@article{Wagner_coherent,
  title = {Exact Coherent Structures and Phase Space Geometry of Preturbulent 2D Active Nematic Channel Flow},
  author = {Wagner, Caleb G. and Norton, Michael M. and Park, Jae Sung and Grover, Piyush},
  journal = {Physical Review Letters},
  volume = {128},
  issue = {2},
  pages = {028003},
  numpages = {7},
  year = {2022},
  month = {Jan},
  publisher = {American Physical Society},
  doi = {10.1103/PhysRevLett.128.028003},
  url = {https://doi.org/10.1103/PhysRevLett.128.028003}
}

@article{ngo2014large,
  title={Large-scale chaos and fluctuations in active nematics},
  author={Ngo, Sandrine and Peshkov, Anton and Aranson, Igor S and Bertin, Eric and Ginelli, Francesco and Chat{\'e}, Hugues},
  journal={Physical Review Letters},
  volume={113},
  url={https://doi.org/10.1103/PhysRevLett.113.038302},
  number={3},
  pages={038302},
  year={2014},
  publisher={APS}
}

@article{decamp2015orientational,
  title={Orientational order of motile defects in active nematics},
  author={DeCamp, Stephen J and Redner, Gabriel S and Baskaran, Aparna and Hagan, Michael F and Dogic, Zvonimir},
  journal={Nature Materials},
  volume={14},
  url={https://doi.org/10.1038/nmat4387},
  number={11},
  pages={1110--1115},
  year={2015},
  publisher={Nature Publishing Group UK London}
}

@book{beris1994thermodynamics,
  title={Thermodynamics of flowing systems: with internal microstructure},
  author={Beris, Antony N and Edwards, Brian J},
  number={36},
  year={1994},
  publisher={Oxford University Press, USA}
}

@article{Mozaffari_spirograph,
  title = {Defect Spirograph: Dynamical Behavior of Defects in Spatially Patterned Active Nematics},
  author = {Mozaffari, Ali and Zhang, Rui and Atzin, Noe and de Pablo, Juan J.},
  journal = {Physical Review Letters},
  volume = {126},
  issue = {22},
  pages = {227801},
  numpages = {6},
  year = {2021},
  month = {Jun},
  publisher = {American Physical Society},
  doi = {10.1103/PhysRevLett.126.227801},
  url = {https://doi.org/10.1103/PhysRevLett.126.227801}
}

@misc{calderer_chevron_2024,
	title = {Chevron patterns in an active nematic liquid crystal film in contact with {Smectic} {A}},
	url = {https://doi.org/10.48550/arXiv.2407.01740},
	urldate = {2024-08-16},
	publisher = {arXiv},
	author = {Calderer, M. Carme and Yao, Lingxing and Zhao, Longhua and Golovaty, Dmitry and Ign\'es-Mullol, Jordi and Sagu\'es, Francesc},
	year = {2024},
	note = {arXiv:2407.01740 [cond-mat]},
}

@incollection{inaba_assembling_2022,
	address = {New York, NY},
	title = {Assembling {Microtubule}-{Based} {Active} {Matter}},
	volume = {2430},
	isbn = {978-1-07-161982-7 978-1-07-161983-4},
	url = {https://doi.org/10.1007/978-1-0716-1983-4_10},
	urldate = {2024-08-16},
	booktitle = {Microtubules},
	publisher = {Springer US},
	author = {Tayar, Alexandra M. and Lemma, Linnea M. and Dogic, Zvonimir},
	editor = {Inaba, Hiroshi},
	year = {2022},
	doi = {10.1007/978-1-0716-1983-4_10},
	note = {Series Title: Methods in Molecular Biology},
	pages = {151--183},
}

@article{memarian_active_2021,
	title = {Active nematic order and dynamic lane formation of microtubules driven by membrane-bound diffusing motors},
	volume = {118},
	issn = {0027-8424, 1091-6490},
	url = {https://doi.org/10.1073/pnas.2117107118},
	doi = {10.1073/pnas.2117107118},
	number = {52},
	urldate = {2024-08-16},
	journal = {Proceedings of the National Academy of Sciences},
	author = {Memarian, Fereshteh L. and Lopes, Joseph D. and Schwarzendahl, Fabian Jan and Athani, Madhuvanthi Guruprasad and Sarpangala, Niranjan and Gopinathan, Ajay and Beller, Daniel A. and Dasbiswas, Kinjal and Hirst, Linda S.},
	year = {2021},
	pages = {e2117107118},
}

@article{ellis2018curvature,
  title={Curvature-induced defect unbinding and dynamics in active nematic toroids},
  author={Ellis, Perry W and Pearce, Daniel JG and Chang, Ya-Wen and Goldsztein, Guillermo and Giomi, Luca and Fernandez-Nieves, Alberto},
  journal={Nature Physics},
  volume={14},
  url={https://doi.org/10.1038/nphys4276},
  number={1},
  pages={85--90},
  year={2018},
  publisher={Nature Publishing Group UK London}
}

@article{PhysRevX.9.041047,
  title = {Hydrodynamics of Active Defects: From Order to Chaos to Defect Ordering},
  author = {Shankar, Suraj and Marchetti, M. Cristina},
  journal = {Physical Review X},
  volume = {9},
  issue = {4},
  pages = {041047},
  numpages = {18},
  year = {2019},
  month = {Dec},
  publisher = {American Physical Society},
  doi = {10.1103/PhysRevX.9.041047},
  url = {https://doi.org/10.1103/PhysRevX.9.041047}
}

@article{thijssen2021submersed,
  title={Submersed micropatterned structures control active nematic flow, topology, and concentration},
  author={Thijssen, Kristian and Khaladj, Dimitrius A and Aghvami, S Ali and Gharbi, Mohamed Amine and Fraden, Seth and Yeomans, Julia M and Hirst, Linda S and Shendruk, Tyler N},
  journal={Proceedings of the National Academy of Sciences},
  volume={118},
  number={38},
  url={https://doi.org/10.1073/pnas.2106038118},
  pages={e2106038118},
  year={2021},
  publisher={National Acad Sciences}
}

@article{PhysRevE.90.062307,
  title = {Active nematic materials with substrate friction},
  author = {Thampi, Sumesh P. and Golestanian, Ramin and Yeomans, Julia M.},
  journal = {Physical Review E},
  volume = {90},
  issue = {6},
  pages = {062307},
  numpages = {6},
  year = {2014},
  month = {Dec},
  publisher = {American Physical Society},
  doi = {10.1103/PhysRevE.90.062307},
  url = {https://doi.org/10.1103/PhysRevE.90.062307}
}

@article{zhang2021spatiotemporal,
  title={Spatiotemporal control of liquid crystal structure and dynamics through activity patterning},
  author={Zhang, Rui and Redford, Steven A and Ruijgrok, Paul V and Kumar, Nitin and Mozaffari, Ali and Zemsky, Sasha and Dinner, Aaron R and Vitelli, Vincenzo and Bryant, Zev and Gardel, Margaret L and others},
  journal={Nature Materials},
  volume={20},
  number={6},
  url={https://doi.org/10.1038/s41563-020-00901-4},
  pages={875--882},
  year={2021},
  publisher={Nature Publishing Group UK London}
}

@article{zarei2023light,
  title={Light-activated microtubule-based two-dimensional active nematic},
  author={Zarei, Zahra and Berezney, John and Hensley, Alexander and Lemma, Linnea and Senbil, Nesrin and Dogic, Zvonimir and Fraden, Seth},
  journal={Soft Matter},
  volume={19},
  number={35},
  url={http://dx.doi.org/10.1039/D3SM00270E},
  pages={6691--6699},
  year={2023},
  publisher={Royal Society of Chemistry}
}

@article{PhysRevLett.132.018301,
  title = {Vortex Lattices in Active Nematics with Periodic Obstacle Arrays},
  author = {Schimming, Cody D. and Reichhardt, C. J. O. and Reichhardt, C.},
  journal = {Physical Review Letters},
  volume = {132},
  issue = {1},
  pages = {018301},
  numpages = {6},
  year = {2024},
  month = {Jan},
  publisher = {American Physical Society},
  doi = {10.1103/PhysRevLett.132.018301},
  url = {https://doi.org/10.1103/PhysRevLett.132.018301}
}

@book{Chung_2002, place={Cambridge}, title={Computational Fluid Dynamics}, publisher={Cambridge University Press}, author={Chung, T. J.}, year={2002}}

@article{upwind,
author = {Courant, Richard and Isaacson, Eugene and Rees, Mina},
title = {On the solution of nonlinear hyperbolic differential equations by finite differences},
journal = {Communications on Pure and Applied Mathematics},
volume = {5},
number = {3},
pages = {243-255},
doi = {https://doi.org/10.1002/cpa.3160050303},
year = {1952}
}

@article{schimming2023friction,
  title={Friction-mediated phase transition in confined active nematics},
  author={Schimming, Cody D and Reichhardt, CJO and Reichhardt, Charles},
  journal={Physical Review E},
  volume={108},
  url={https://doi.org/10.1103/PhysRevE.108.L012602},
  number={1},
  pages={L012602},
  year={2023},
  publisher={APS}
}

@article{mirantsev2021behavior,
  title={Behavior of chiral active nematics confined to nanoscopic circular region},
  author={Mirantsev, LV},
  journal={The European Physical Journal E},
  volume={44},
  url={https://doi.org/10.1140/epje/s10189-021-00120-y},
  number={9},
  pages={112},
  year={2021},
  publisher={Springer}
}

@article{joshi2023disks,
  title={From disks to channels: dynamics of active nematics confined to an annulus},
  author={Joshi, Chaitanya and Zarei, Zahra and Norton, Michael M and Fraden, Seth and Baskaran, Aparna and Hagan, Michael F},
  journal={Soft Matter},
  volume={19},
  url={DOI	https://doi.org/10.1039/D3SM00477E},
  number={29},
  pages={5630--5640},
  year={2023},
  publisher={Royal Society of Chemistry}
}

@article{velez2024active,
author = {Ignasi Vélez-Cerón  and Rodrigo C. V. Coelho  and Pau Guillamat  and Marc Vergés-Vilarrubia  and Margarida Telo da Gama  and Francesc Sagués  and Jordi Ignés-Mullol },
title = {Active nematic pumps},
journal = {Proceedings of the National Academy of Sciences},
volume = {122},
number = {46},
pages = {e2427103122},
year = {2025},
doi = {10.1073/pnas.2427103122},
URL = {https://doi.org/10.1073/pnas.2427103122},
}

@article{roberts2019ensemble,
  title={Ensemble-based topological entropy calculation ({E}-tec)},
  author={Roberts, Eric and Sindi, Suzanne and Smith, Spencer A.\ and Mitchell, Kevin A},
  journal={Chaos: An Interdisciplinary Journal of Nonlinear Science},
  volume={29},
  url={https://doi.org/10.1063/1.5045060},
  number={1},
  year={2019},
  publisher={AIP Publishing}
}

@misc{weisstein2003epicycloid,
  title={Epicycloid},
  author={Weisstein, Eric W},
  howpublished={\url{https://mathworld.wolfram.com/Epicycloid.html}},
  year={2003},
}

@article{shankar2024design,
  title={Design rules for controlling active topological defects},
  author={Shankar, Suraj and Scharrer, Luca VD and Bowick, Mark J and Marchetti, M Cristina},
  journal={Proceedings of the National Academy of Sciences},
  volume={121},
  number={21},
  url={https://doi.org/10.1073/pnas.2400933121},
  pages={e2400933121},
  year={2024},
  publisher={National Acad Sciences}
}

@article{schimming2024vortex,
  title={Vortex lattices in active nematics with periodic obstacle arrays},
  author={Schimming, Cody D and Reichhardt, CJO and Reichhardt, Charles},
  journal={Physical Review Letters},
  volume={132},
  url={https://doi.org/10.1103/PhysRevLett.132.018301},
  number={1},
  pages={018301},
  year={2024},
  publisher={APS}
}

@article{turiv2020polar,
  title={Polar jets of swimming bacteria condensed by a patterned liquid crystal},
  author={Turiv, Taras and Koizumi, Runa and Thijssen, Kristian and Genkin, Mikhail M and Yu, Hao and Peng, Chenhui and Wei, Qi-Huo and Yeomans, Julia M and Aranson, Igor S and Doostmohammadi, Amin and others},
  journal={Nature Physics},
  volume={16},
  url={https://doi.org/10.1038/s41567-020-0793-0},
  number={4},
  pages={481--487},
  year={2020},
  publisher={Nature Publishing Group UK London}
}

@article{norton2018insensitivity,
  title={Insensitivity of active nematic liquid crystal dynamics to topological constraints},
  author={Norton, Michael M and Baskaran, Arvind and Opathalage, Achini and Langeslay, Blake and Fraden, Seth and Baskaran, Aparna and Hagan, Michael F},
  journal={Physical Review E},
  volume={97},
  number={1},
  url={https://doi.org/10.1103/PhysRevE.97.012702},
  pages={012702},
  year={2018},
  publisher={APS}
}

@article{santo_dpd_thermo,
title = {Dissipative particle dynamics simulations in colloid and Interface science: a review},
journal = {Advances in Colloid and Interface Science},
volume = {298},
pages = {102545},
year = {2021},
issn = {0001-8686},
doi = {https://doi.org/10.1016/j.cis.2021.102545},
author = {Kolattukudy P. Santo and Alexander V. Neimark}
}

@misc{flowSolverGithub,
author = {Brandon Klein and
Alejandro J.\ Soto Franco and
Md Mainul Hasan Sabbir and
Matthew J.~Deutsch and
Ross Kliegman and Robin L.~B.~Selinger and Kevin A. Mitchell and Daniel A.\ Beller}
,
title = {Spontaneous-Optimal-Mixing},
year = {2025},
howpublished = {\url {https://github.com/Brandonkl/Spontaneous-Optimal-Mixing}},
doi = {https://doi.org/10.5281/zenodo.15733450}
}

@article{ben2022disordered,
  title={Disordered boundaries destroy bulk phase separation in scalar active matter},
  author={Ben Dor, Ydan and Ro, Sunghan and Kafri, Yariv and Kardar, Mehran and Tailleur, Julien},
  journal={Physical Review E},
  volume={105},
  number={4},
  pages={044603},
  year={2022},
  publisher={APS},
doi = {https://doi.org/10.1103/PhysRevE.105.044603}
}

@article{velez2024probing,
  title={Probing active nematics with in situ microfabricated elastic inclusions},
  author={V{\'e}lez-Cer{\'o}n, Ignasi and Guillamat, Pau and Sagu{\'e}s, Francesc and Ign{\'e}s-Mullol, Jordi},
  journal={Proceedings of the National Academy of Sciences},
  volume={121},
  number={11},
  pages={e2312494121},
  year={2024},
  publisher={National Academy of Sciences},
doi = {https://doi.org/10.1073/pnas.2312494121}
}

\clearpage

\begin{center}
    \Large \textbf{Supporting Information}
\end{center}
\appendix
\section{Numerical methods}

\subsection{Nematohydrodynamics \label{sec: methods-nematohydrodynamics}}

\subsubsection{Pressure field}

In the incompressible Navier-Stokes equations 
12, 13, the pressure field, $p$, plays the exclusive role of maintaining Eq. 13. 
This is achieved in our numerical implementation by taking the divergence of Eq. 12 
and solving the standard pressure-Poisson scheme \cite{Chung_2002}, keeping terms up to second order in derivatives of $\mathbf{u}$:
\begin{equation}
    \nabla^2 p = -\nabla \cdot ({\bf u} \cdot \nabla){\bf u} + \nabla \cdot \frac{1}{\rho}{\bf F} - \nabla \cdot \partial_{t} {\bf u} \rvert_{t}
    \label{eq:pressure-poisson}
    \tag{SI.1}
\end{equation}
with a Laplacian stencil of $p$, such that $\nabla \cdot \partial_{t} {\bf u}\rvert_{t+\delta t }$ = 0. 

 We integrate equations 8
 and 10
 forward in time with a time step of $\delta t = 1\mathrm{e}{-4}$ using the Euler method \cite{flowSolverGithub}. Advection terms for $\bf u$ and $Q_{ij}$ are calculated using an upwind scheme which computes advection coming from the direction of the local flow field  \cite{upwind}.

\subsubsection*{Boundary conditions}

To simulate strong tangential anchoring on an arbitrarily curved boundary with local unit tangent $\hat \tau$, we apply Dirichlet conditions on the nematic domain $\Omega$, with boundary $\partial\Omega$. Specifically,
\begin{equation}
    \begin{pmatrix}
        Q_{xx} \\
        Q_{xy}
    \end{pmatrix}\bigg\rvert_{\partial \Omega} ({\bf r}) = 
       S\begin{pmatrix}
        \tau_x^2 - 1/2 \\
        \tau_x\tau_y
    \end{pmatrix}.
    \tag{SI.2}
\end{equation}

In order to provide the force from such a boundary that would result in this anchoring, or equivalently that $\partial_t Q_{ij} = 0$, we set the molecular field along $\partial \Omega$ based on equation 8
as 
\begin{equation}
    H_{ij}\bigg\rvert_{\partial \Omega} = \gamma[u_k\partial_kQ_{ij} - \chi SE_{ij} + [\omega, Q] + 2\mathrm{Tr}[QE]Q_{ij}].
    \tag{SI.3}
\end{equation}
For the flow velocity field, we use a no-slip boundary condition, ${\bf u}\big\rvert_{\partial \Omega} = 0$. Equation \ref{eq:pressure-poisson} is an instance of Poisson's equation, which has a unique solution so long as Neumann or Dirichlet boundary conditions are defined. To do this, we consider that there are no outflows or inflows along the boundary,  $\hat \nu \cdot {\bf u}\big\rvert_{\partial \Omega} = 0$, and derive a Neumann boundary condition using $\hat \nu \cdot \partial_t {\bf u}\big\rvert_{\partial \Omega} = 0$, with $\hat \nu$ being the outward-pointing unit normal to the boundary and with
\begin{align}
     \hat \nu \cdot \partial_t {\bf u}\big\rvert_{\partial \Omega}  & = \hat \nu \cdot \left[-({\bf u} \cdot \nabla){\bf u} + \eta \nabla^2 {\bf u} + \frac{1}{\rho}{\bf F} - \frac{1}{\rho}\nabla { p}\right]\bigg\rvert_{\partial \Omega}.
     \tag{SI.4}
\end{align}
Upon applying the no-flux condition at the boundary,
$\left. \hat \nu \cdot \mathbf{u} \right|_{\partial \Omega} = 0$, and the corollary that the tangential derivative of the normal component of velocity must vanish, $\left. \partial_\tau (\hat \nu \cdot \mathbf{u}) =0 \right|_{\partial\Omega} $, we obtain the following Neumann condition for the normal derivative of the pressure at the boundary:
\begin{align}
  \partial_\nu p\big\rvert_{\partial \Omega} & = \left(\rho \eta \nabla^2 { u_\nu} + {F_\nu} \right)\rvert_{\partial\Omega}.
  \tag{SI.5}
\end{align}

Note that $u_\nu \big\rvert_{\partial \Omega} = 0$ even if there is slipping, which necessarily occurs along $\hat \tau$. Importantly, our scheme is generalizable to $\mathcal{C}^1$-continuous curves.

For compatibility with our finite differencing scheme, we need to regularize the cusps of the epicycloids into  $\mathcal{C}^1$-continuous curves, which also enables us to model the smoothness of the "cusp" in the experiments of Ref.~\cite{mitchell_cardioid} at the scale of the microtubule bundles. For this purpose, we use as our family of boundary surfaces the epitrochoids defined parametrically by
\begin{equation} \label{eq:epitrochoid}
    \begin{aligned}  
        x(u) = \frac{r}{2q}\left[(2q-1)\cos(u) + d\cos((2q-1)u)\right]\\
        y(u) = \frac{r}{2q}\left[(2q-1)\sin(u) + d\sin((2q-1)u)\right]\\
        u\in [0,2\pi),
    \end{aligned}
    \tag{SI.6}
\end{equation}
where the new parameter $d$ varies between $0$ and $1$ continuously interpolates between a circle and the epicycloids. These represent paths tracing a point on a circle of radius $r$ at distance $r \cdot d$ from its center as it rolls on the circumference of a circle $R$. We use $d = 0.99$ to approximate the epicycloids near their sharp limit.

\subsection{Time averaged vorticity on epicycloid simulations}
Our theoretical framework predicts that the number of gyres in the time averaged vorticity only depends on boundary topology, not geometry. We show that the epicycloid simulations performed agree with our analytical prediction, by showing the time averaged vorticity as a function of space for the braiding regimes discussed in Figure 7 of the main text.

\begin{figure*}[h!]
\centering
\includegraphics[width=\textwidth]{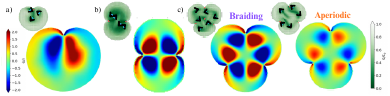}\caption{(a), (b), and (c) show the time averaged vorticity after $1.5 \times 10^6$ time-steps for the braiding and aperiodic regimes in the cardioid, nephroid, and trefoiloid. Insets show the corresponding snapshots from Figure 7 (a), (c), and (e).}
\end{figure*}

\subsection{Agent-based model \label{sec: methods-agent-based}} 
The three-dimensional coarse-grained dynamical simulations represent active microtubules as bead-spring chain filaments, confined within a volume defined by a cardioid-shaped area with a small thickness, $L = 3.2\sigma$, in the vertical direction.  Interactions between beads are represented by a short-range, repulsive, Weeks-Chandler-Anderson interaction with an additional short-range attractive depletion force:
\begin{equation} \label{eq:wca}
U(r)= 
	\begin{cases}
		4 \varepsilon \left [ \left ( \frac{\sigma}{r} \right )^{12} - \left ( \frac{\sigma}{r} \right )^6\right] + f_\text{dep} r, & \text{if $r < 2^{\frac{1}{6}}\sigma$}, \\
		0, & \text{if $r \geq 2^{\frac{1}{6}}\sigma$}.
	\end{cases}
    \tag{SI.7}
\end{equation}
The bead-bead interaction has parameters of $\varepsilon = 0.5$, $\sigma = 1.0$, and the additional depletion force is ${\bf f}_\text{dep} = 0.25$. Bonded interactions between adjacent beads on each chain are represented as linear elastic springs with potential:
\begin{equation} \label{eq:ubond}
	U_\text{bond}(r) = \frac{k_1}{2} (r - l_0)^2
    \tag{SI.8}
\end{equation}
where $l_0$ is the equilibrium length of the spring and $k_1$ is the spring constant. Bond-bending terms are represented by second- and third-neighbor linear spring interactions with spring constants $k_2$ and $k_3$ and equilibrium lengths $2l_0$ and $3l_0$:
\begin{equation} \label{eq:ubend}
	U_\text{bend} = \frac{k_2}{2} (|{\bf r}_{i+2}-{\bf r}_{i}| - 2l_0)^2 + \frac{k_3}{2} (|{\bf r}_{i+3} -{\bf r}_{i}|- 3l_0)^2 
    \tag{SI.9}
\end{equation}
In Eq.~\ref{eq:ubond} and Eq.~\ref{eq:ubend} equilibrium length, $l_0 = 0.8$, bond spring constant, $k_1 = 57.1464$, and bend spring constants $k_2 = k_3 = 50k_1$. The equations of motion for the beads, with mass $m=1$, are integrated using the Velocity Verlet algorithm.

Extensile activity is modeled via a pairwise active force between beads $i$ and $j$ on adjacent filaments with anti-polar alignment, of the form:

\begin{equation} \label{eq:factive}
	{\bf F}_i = \alpha \frac{\frac{1}{2}(\hat{t}_i - \hat{t}_j)}{|{\bf r}_{ij}|}, \qquad {\bf F}_j = -{\bf F}_i 
,\tag{SI.10}\end{equation}
where $\hat{t}$ is the tangent vector of the adjacent filaments at the positions of beads $i$ and $j$ and activity parameter, $\alpha = 0.06$. We note that this active filament activity parameter, $\alpha$, does not map to the nematohydrodynamic activity, $\zeta$. We add a constant short-range attractive force, ${\bf f}_\text{attract}$, to the pairwise active force to represent the attractive pull of kinesin motor proteins on two adjacent filaments undergoing shear. Activity is only applied to adjacent filaments if $\hat{t}_i \cdot \hat{t}_j \leq 0.5$, resulting in extensile activity through inter-filament shear. Our coarse-grained model assumes that there is a uniform, high density, of kinesin motor proteins and adenosine triphosphate chemical energy (ATP). The model thus does not show the characteristic slowing of extensile shear as ATP density drops with time as seen in experiment~\cite{memarian_active_2021}.

To enable long-range hydrodynamic interactions, we introduce a coarse-grained two-dimensional fluid layer located below the filament volume and coupled to the filaments.  This novel underlying fluid layer is introduced for two effects: to thermostat the active matter, and to provide long-range hydrodynamic interactions across areas with low active filament density. Fluid-fluid particle interactions are governed by a short-range, repulsive, Weeks-Chandler-Anderson potential similar to Eq. \ref{eq:wca}. Fluid particles are thermostatted by a pairwise dissipative particle dynamics thermostat ~\cite{santo_dpd_thermo}. Interactions between fluid particles and active particles are represented by a Lennard-Jones interaction with interaction strength mediated by an artificial distance offset between the fluid layer and the active layer. This interaction puts all active particles in the simulation volume in contact with the fluid thermostat.

Arbitrarily shaped boundary conditions can be imposed on both active and fluid particles in the system by generating a wall of immobile boundary particles which have an exclusively repulsive soft-sphere potential interaction. For the single-cusp cardioid system the boundary particles were laid out with the epicycloid equations for $q = 3/2$,
where $r = 123\sigma$ is the characteristic radius of the cardioid and $u$ is calculated to produce equally-spaced boundary particles along the perimeter of the cardioid. The spacing between boundary particles is calculated to give a uniform repulsive force along the boundary and to be small enough to effectively confine the fluid and active particles.

\section{Analytical methods \label{sec: Burau}}
\subsection{The Artin braid group and topological entropy}
The Artin braid groups are closed under a product operation. The Burau representation is given by \cite{thiffeault_measuring_2005}
\begin{align} \label{Burau}
    (\sigma_i)_{kl} = \delta_{kl} + \delta_{i-1,k}\delta_{il} - \delta_{i+1,k}\delta_{il}, \nonumber \\
    (\sigma_i^{-1})_{kl} = \delta_{kl} - \delta_{i-1,k}\delta_{il} + \delta_{i+1,k}\delta_{il}.
\end{align}
and consists of $2(n-1)$ matrices of size $(n-1)\times(n-1)$ where the empty product and group identity is mapped to the $n-1$ dimensional identity matrix. This matrix representation is constructed to maintain the Artin group relations: $[\sigma_i, \sigma_j] = 0 \text{ if } |i - j| > 1$, and $\sigma_i \sigma_{i+1} \sigma_i = \sigma_{i+1} \sigma_i \sigma_{i+1}$. Importantly, this means that a braidword can be represented as a matrix product, and that a periodic steady state corresponds to the application of $\beta^{n_c}$ for $n_c$ cycles. In the large-$n_c$  limit, the matrix product $\beta^{n_c}$ in its eigenbasis is dominated by its largest-magnitude eigenvalue $b_{\mathrm{max}}^{n_c}$, where $b_{\mathrm{max}}$ is the largest eigenvalue of $\beta$. Because the defects, as stirring rods, drag the fluid with them, the minimal stretching of material contours required to accommodate the described defect braiding grows with $n_c$ as $b_{\mathrm{max}}^{n_c}$. The topological entropy therefore grows linearly with $n_c$, as $h\cdot t \sim \log(|b_{\mathrm{max}}^{n_c}|) = n_c \log(|b_{\mathrm{max}}|)$. Since $n_c$ is proportional to time for periodic braiding, $h$ is proportional to $ \log(|b_{\mathrm{max}}|)$.  Note that, if the motion is periodic, $h$ is independent of the projection used.

\subsection{Burau representation of \texorpdfstring{$\mathbf{B}_3$ and $\mathbf{B}_4$}{B3 and B4}}
The elements of $\mathbf{B}_3$ in the Burau representation are
\begin{equation}
\begin{aligned}
    \sigma_1 = \begin{pmatrix}1 &1 \\ 0& 1\end{pmatrix}, \; \sigma_1^{-1} = \begin{pmatrix}1 &-1 \\ 0& 1\end{pmatrix}, \;
    \\ \sigma_2 = \begin{pmatrix}1 &0 \\ -1& 1\end{pmatrix}, \; \sigma_2^{-1} = \begin{pmatrix}1& 0 \\ 1 &1\end{pmatrix}.
    \end{aligned}
    \tag{SI.11}
\end{equation}

An iteration of the golden braid then looks like
\begin{equation}
\begin{aligned}
 \beta_\text{golden} = \sigma_2^{-1}\sigma_1 = \begin{pmatrix}1 &0 \\ 1& 1\end{pmatrix} \begin{pmatrix}1& 1 \\ 0 &1\end{pmatrix} = \begin{pmatrix}1& 1 \\ 1 &2\end{pmatrix}.
\end{aligned}
\tag{SI.12}
\end{equation}

The eigenvalues of this matrix are $\frac{3 + \sqrt 5}{2} = 1 + \phi_0 = \phi_0^2$, and $\frac{3 - \sqrt 5}{2} = 1 - (\phi_0 - 1) = (\phi_0 - 1)^2$, where $\phi_0 = \frac{1 +\sqrt 5}{2}$ is the golden ratio. Another method to see the rate of stretching is to consider the action of this braidword on an arbitrary vector $\begin{pmatrix} x\\y\end{pmatrix},$ which gives

\begin{equation}
\begin{aligned}
 \begin{pmatrix} x'\\y'\end{pmatrix} = \beta_\text{golden} \begin{pmatrix} x\\y\end{pmatrix}= \begin{pmatrix}1& 1 \\ 1 &2\end{pmatrix} \begin{pmatrix} x\\y\end{pmatrix} \\= \begin{pmatrix}
     x + y \\x + 2y
 \end{pmatrix}.
 \end{aligned}
 \tag{SI.13}
\end{equation}

Let $F_k$ be the $k^\text{th}$ Fibonacci number. It holds that if $x = F_{n-2}$, and $y = F_{n-1}$, then 
\begin{equation}
x' = F_{n-1} + F_{n-2} = F_{n}, \tag{SI.14}
\end{equation}
and 
\begin{equation}
y'= 2F_{n-1} + F_{n-2} = F_{n} + F_{n-1} = F_{n+1}. \tag{SI.15}
\end{equation}

Thus, by induction, powers of $\beta_\text{golden}$ produce the Fibonacci sequence.

The elements of $\mathbf{B}_4$ are given as
\begin{equation}
\begin{aligned}
\sigma_1&=
\begin{pmatrix}
  1 & 0& 0 \\
  -1& 1& 0 \\
  0 & 0& 1
\end{pmatrix}, \;
\sigma_1^{-1}&=
\begin{pmatrix}
  1 & 0& 0 \\
  1 & 1& 0 \\
  0 & 0& 1
\end{pmatrix}, \;\\
\sigma_2&=
\begin{pmatrix}
  1 & 1& 0 \\
  0 & 1& 0 \\
  0 &-1& 1
\end{pmatrix},
\sigma_2^{-1}&=
\begin{pmatrix}
  1 &-1& 0 \\
  0 & 1& 0 \\
  0 &1& 1
\end{pmatrix}, \;\\
\sigma_3&=
\begin{pmatrix}
  1 & 0& 0 \\
  0 & 1& 1 \\
  0 & 0& 1
\end{pmatrix}, \;
\sigma_3^{-1}&=
\begin{pmatrix}
  1 & 0& 0 \\
  0 & 1&-1 \\
  0 & 0& 1
\end{pmatrix}.
\end{aligned}
\tag{SI.16}
\end{equation}

An iteration of the silver braid, $\sigma_3\sigma_1\sigma_2\sigma_3^{-1}\sigma_1^{-1}\sigma_2^{-1}$, then looks like 
\begin{equation}
\begin{aligned}
 \beta_\text{silver} = \sigma_3\sigma_1\sigma_2\sigma_3^{-1}\sigma_1^{-1}\sigma_2^{-1} \\ 
 =
 \begin{pmatrix}
  1 & 0& 0 \\
  0 & 1& 1 \\
  0 & 0& 1
\end{pmatrix}
\begin{pmatrix}
  1 & 0& 0 \\
  -1& 1& 0 \\
  0 & 0& 1
\end{pmatrix}\\
\begin{pmatrix}
  1 & 1& 0 \\
  0 & 1& 0 \\
  0 &-1& 1
\end{pmatrix}
\begin{pmatrix}
  1 & 0& 0 \\
  0 & 1&-1 \\
  0 & 0& 1
\end{pmatrix}\\
\begin{pmatrix}
  1 & 0& 0 \\
  1 & 1& 0 \\
  0 & 0& 1
\end{pmatrix}
\begin{pmatrix}
  1 &-1& 0 \\
  0 & 1& 0 \\
  0 &1& 1
\end{pmatrix} \\
 = \begin{pmatrix} 
 2& -2& -1\\
 -2&  3&  2\\
 -1&  2&  2\\\end{pmatrix}.
\end{aligned}
\tag{SI.17}
\end{equation}

The eigenvalues of this matrix are $3 + 2\sqrt2 = 1 + 2\phi_1 = \phi_1^2$, $3 - 2\sqrt2 = 1 - 2(\phi_1 - 2) = (\phi_1 - 2)^2$, and $1 = \phi_1(\phi_1 - 2)$, where $\phi_1 = 1 + \sqrt{2}$ is the silver ratio. By definition both the golden and silver ratios obey the conjugate and identity relations of the metallic ratios:
\begin{equation}
\begin{aligned}
1 + k\phi_{k-1} &= \phi_{k-1}^2, \\
1 - k(\phi_{k-1} - k) &= (\phi_{k-1} - k)^2,
\end{aligned}
\tag{SI.18}
\end{equation}
and thus,
\begin{equation}
\phi_{k-1}(\phi_{k-1} - k) = 1.
\tag{SI.19}
\end{equation}

\section{Supplemental video captions}
\begin{enumerate}[label=Movie S\arabic*]

\item{Simulation over $3.35\times 10^5$ time-steps of a $100 \times 100$ simulation of an active nematic confined to a disk with fixed tangential anchoring. Defect trajectories (blue and green) are shown in the disk on the left. Defect worldlines are displayed on the right. The projection axis denoted ``X" is the horizontal axis of the disk.}

\item{(0-50 seconds) Simulation over $7.5\times 10^5$ time-steps of a $100 \times 100$  simulation of an active nematic confined to a disk with fixed $q=3/2$ anchoring. Defect trajectories (blue, green, and purple) are shown in the disk on the left. Defect worldlines are displayed on the right. The projection axis denoted ``X" is the horizontal axis of the disk. (50-66 seconds) Line stretching depicts an advected contour (blue) undergoing exponential stretching due to defect mixing (red). The director outside the circular domain is arbitrarily defined as horizontal and not simulated.}

\item{(0-33 seconds) Simulation over $5.0\times10^5$ time-steps of a $100 \times 100$  simulation of an active nematic confined to a disk with fixed q=4/2 anchoring. Defect trajectories (blue, green, and purple) are shown in the disk on the left. Defect worldlines are displayed on the right. The projection axis denoted “X” is the horizontal axis of the disk. (50-49 seconds) Line stretching depicts an advected contour (blue) undergoing exponential stretching due to defect mixing (red). The director outside the circular domain is arbitrarily defined as horizontal and not simulated.} 

\item{Simulation over $7.5\times 10^5$ time-steps of a $100 \times 100$ simulation of an active nematic confined to a disk with fixed $q=3/2$ anchoring. The top three graphs show the instantaneous vorticity, the running time-averaged vorticity, and the running standard deviation of the vorticity. The bottom three graphs show the instantaneous $\mathcal{Q}$-criterion with $\mathcal{Q} = 0$ isolines shown in black, the running time-averaged $\mathcal{Q}$-criterion, and the running standard deviation of the $\mathcal{Q}$-criterion.}

\item{Simulation over $5\times 10^5$ time-steps of a $100 \times 100$ simulation of an active nematic confined to a disk with fixed $q=4/2$ anchoring. The top three graphs show the instantaneous vorticity, the running time-averaged vorticity, and the running standard deviation of the vorticity. The bottom three graphs show the instantaneous $\mathcal{Q}$-criterion with $\mathcal{Q} = 0$ isolines shown in black, the running time-averaged $\mathcal{Q}$-criterion, and the running standard deviation of the $\mathcal{Q}$-criterion.}

\item{Simulation over $1.3\times10^6$ time-steps of a $100 \times 100$ simulation of an active nematic confined to a disk with fixed $q=5/2$ anchoring. The top three graphs show the instantaneous vorticity, the running time-averaged vorticity, and the running standard deviation of the vorticity. The bottom three graphs show the instantaneous $\mathcal{Q}$-criterion with $\mathcal{Q} = 0$ isolines shown in black, the running time-averaged $\mathcal{Q}$-criterion, and the running standard deviation of the $\mathcal{Q}$-criterion.}

\item{Representative videos of the active phases seen in Figure 
7b. All simulations are performed on a $200 \times 200$ lattice for $1.5\times 10^6$ time-steps. $(\ell_a, \ell_c)$ values for shown simulations are $(0.0069, 0.0625)$ for turbulent, $(0.0556, 0.0834)$ for arrested, $(0.0417, 0.0486)$ for interrupted (golden), and $(0.0139, 0.0903)$ for golden braid.}

\item{Representative videos of the active phases seen in Figure 
7d. All simulations are performed on a $100 \times 100$ lattice for $1.5\times 10^6$ time-steps. $(\ell_a, \ell_c)$ values for shown simulations are $(0.0131, 0.0131)$ for turbulent, $(0.0262, 0.0131)$ for arrested, $(0.0262, 0.0654)$ for interrupted (silver), $(0.0196, 0.0393)$ for mixed interrupted, $(0.0131, 0.1309)$ for melted, and $(0.0131, 0.1178)$ for silver braid.}

\item{Representative videos of the active phases seen in Figure 
7e. All simulations are performed on a $200 \times 200$ lattice for $1.5\times 10^6$ time-steps. $(\ell_a, \ell_c)$ values for shown simulations are $(0.0064, 0.0128)$ for turbulent, $(0.0511, 0.0766)$ for arrested, $(0.0192, 0.0766)$ for aperiodic, $(0.0128, 0.0639)$, for interrupted (silver), and $(0.0128, 0.0766)$ for silver braid.}

\item{Agent-based simulation in cardioid confinement of 64,000 active particles comprising 800 filaments, each with 80 beads, and an additional 3068 fluid particles. The left shows bead-chain filaments with green indicating CCW polar orientation and orange indicating CW polar orientation with respect to the origin. The right shows locally averaged director field and defect trajectories in blue, green, and red, performing a golden braid cycle.}

\end{enumerate}

\end{document}